\documentclass[12pt, leqno]{article}
\usepackage{amsmath,setspace,multirow,lineno}
\usepackage[font=small,labelfont=bf,singlelinecheck=off]{caption}
\usepackage[top=1.25in, bottom=1.25in, left=1.1in, right=1.1in]{geometry}

\usepackage[round]{natbib}
\usepackage{color,soul}

\DeclareCaptionStyle{italic}[justification=centering]{labelfont={bf},textfont={it},labelsep=colon}

\usepackage{graphicx,psfrag,epsf,textcomp,epstopdf, amsthm, paralist}
\usepackage{enumerate, dsfont, alltt, verbatim, algorithm, algpseudocode, algorithmicx}
\usepackage{url,xcolor}
\usepackage{booktabs, subfig, bm, paralist,mathpazo,tikz,longtable,microtype, authblk}
\usepackage[utf8]{inputenc}

\usepackage[pdftex,colorlinks=true]{hyperref}
\definecolor{darkblue}{rgb}{0,0,.6}
\hypersetup{citecolor=darkblue,linkcolor=darkblue,urlcolor=darkblue}
\definecolor{DarkRed}{rgb}{.7,0,.4}

\usepackage{comment,orcidlink}

\newcommand{\blind}{0}

\newcommand{\X}{\mathcal{X}}

\newcommand{\Z}{\mathcal{Z}}

\newcommand{\Rlogo}{\protect\includegraphics[height=1.8ex,keepaspectratio]{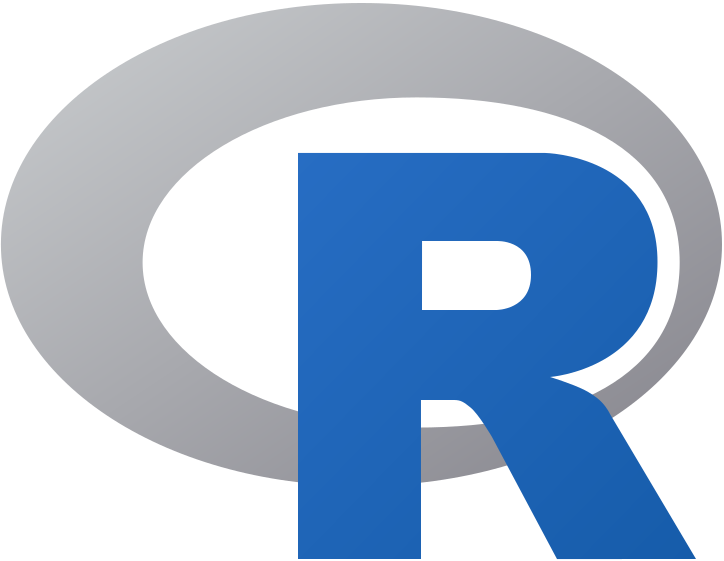}}

\graphicspath{{plots/}}
\DeclareMathOperator*{\argmin}{\arg\!\min}

\newsavebox\CBox

\newtheorem{@definition}{\sc Definition}[section]

\newtheorem{theorem}{\sc Theorem}[section]

\renewcommand\X{\mathcal{X}}

\date{}

\begin{document}

\def\spacingset#1{\renewcommand{\baselinestretch}{#1}\small\normalsize} \spacingset{1}

\if0\blind
{
\title{\bf Spatial Scalar-on-Function Quantile Regression Model}}
\author[1]{\normalsize Muge Mutis \orcidlink{0000-0002-9801-4835}}
\author[2]{\normalsize Ufuk Beyaztas \orcidlink{0000-0002-5208-4950}}
\author[1]{Filiz Karaman \orcidlink{0000-0002-8491-674X}}
\author[3]{\normalsize Han Lin Shang \orcidlink{0000-0003-1769-6430}\footnote{Correspondence: Department of Actuarial Studies and Business Analytics, Macquarie University, NSW 2109, Australia; Telephone: +61(2) 9850 4689; Email: hanlin.shang@mq.edu.au}}

\affil[1]{\normalsize Department of Statistics, Yildiz Technical University, Turkey}
\affil[2]{\normalsize Department of Statistics, Marmara University, Turkey}
\affil[3]{\normalsize Department of Actuarial Studies and Business Analytics, Macquarie University, Australia}

\maketitle
\fi

\if1\blind
{
\title{\bf Spatial Scalar-on-Function Quantile Regression Model}
\author{}
} 
\fi

\maketitle

\begin{abstract}
This paper introduces a novel spatial scalar-on-function quantile regression model that extends classical scalar-on-function models to account for spatial dependence and heterogeneous conditional distributions. The proposed model incorporates spatial autocorrelation through a spatially lagged response and characterizes the entire conditional distribution of a scalar outcome given a functional predictor. To address the endogeneity induced by the spatial lag term, we develop two robust estimation procedures based on instrumental variable strategies. $\sqrt{n}$-consistency and asymptotic normality of the proposed estimators are established under mild regularity conditions. We demonstrate through extensive Monte Carlo simulations that the proposed estimators outperform existing mean-based and robust alternatives, particularly in settings with strong spatial dependence and outlier contamination. We apply our method to high-resolution environmental data from the Lombardy region in Italy, using daily ozone trajectories to predict daily mean particulate matter with a diameter of less than 2.5 micrometers concentrations. The empirical results confirm the superiority of our approach in predictive accuracy, robustness, and interpretability across various quantile levels. Our method has been implemented in the \texttt{ssofqrm} \Rlogo \ package.
\end{abstract}


\newpage
%\spacingset{1.65} 
\doublespacing

\section{Introduction} \label{sec1}

The analysis of data observed over continuous domains, such as curves and surfaces, benefits significantly from treating these observations as holistic functional entities rather than discrete measurements. This functional perspective enables the exploitation of the data's inherent smoothness and structural dependencies, leading to more efficient and interpretable statistical methodologies. Consequently, functional data analysis (FDA) has emerged as a powerful framework, wherein entire functions serve as the primary units of analysis. The increasing prominence of the FDA has driven the development of various statistical techniques tailored to such data structures. For a comprehensive discussion of the theoretical foundations and applied methodologies in FDA, we refer readers to \cite{Ramsay1991}, \cite{RamsaySilverman2006}, \cite{Horvath2012}, and \cite{Kokoszka2017}, to name only a few.

Among the numerous models within the FDA, the scalar-on-function linear regression model (SoFLRM) has garnered considerable attention. This model is specifically designed to explore the association between a functional predictor, represented as a random curve, and a scalar response. By leveraging the rich information embedded in functional regressors, SoFLRM facilitates both inference and prediction in various scientific and applied contexts.

Let $\{Y_{i}, \X_{i}(u)\}$, $i \in \{1, \ldots, n\}$, denote a random sample drawn from a population characterized by the pair $(Y, \X)$. In this framework, the response variable $Y \in \mathbb{R}$ is scalar, whereas the predictor $\X = \{\X(u) \}_{u \in \mathcal{I}}$ represents a functional regressor belonging to the separable and infinite-dimensional Hilbert space $\mathcal{L}^2(\mathcal{I})$, where $\mathcal{I}$ is a bounded and closed interval. For simplicity, we assume that both the response and the regressor are mean-zero processes, i.e., $\mathbb{E}(Y) = \mathbb{E} \{ \X(u) \} = 0$. Under these assumptions, the SoFLRM is defined as follows:
\begin{equation}\label{eq1}
\mathbb{E} \{Y \vert \X(u)\} = \int_{\mathcal{I}} \X(u) \beta(u) \, du,
\end{equation}
where $\mathbb{E} \{ Y \vert \X(u) \}$ denotes the conditional expectation of the scalar response given the functional predictor $\X(u)$ and $\beta(u) \in \mathcal{L}^2(\mathcal{I})$ is an unknown regression coefficient function.

The SoFLRM formulation in~\eqref{eq1} has inspired a vast body of literature, leading to various extensions, including nonlinear and nonparametric adaptations that further refine the role of the functional regressor $\X(u)$ in regression models \citep[see, e.g.,][for a detailed discussion]{Morris15, Reiss2017}. A fundamental assumption underlying these models is the independence of sampled units. However, data often exhibit spatial dependence in numerous real-world applications, spanning fields such as economics, sociology, ecology, and epidemiology. This dependence arises when observations are correlated due to their geographical proximity, implying that conventional SoFLRM approaches and extensions may fail to capture the inherent spatial interactions in such datasets. For instance, crime rates in a given city are typically influenced by those in adjacent cities, and the price of a land parcel is strongly associated with the prices of neighboring parcels. The inability of classical models such as SoFLRM to account for such spatial dependencies underscores the necessity of developing more sophisticated methodologies capable of incorporating spatial structures into functional regression frameworks.

When the predictor is scalar rather than functional, spatial dependence can be incorporated into a linear regression model through two principal approaches: introducing a spatially lagged dependent variable as an additional regressor or modifying the error structure to account for spatial autocorrelation. The first approach, known as the spatial autoregressive model, is particularly useful when the primary objective is to evaluate the presence and magnitude of spatial dependence. In contrast, the second approach, referred to as the spatial error model, is designed to correct for potential biases introduced by spatial autocorrelation, regardless of whether the model explicitly includes a spatial component \citep[see, e.g.,][for further details]{Anselin2003, Lesage2009}.

This study focuses on the spatial autoregressive model, which provides a principled framework for incorporating spatial dependence into the regression structure. When the observations $\{Y_{i}, \X_{i}(u) \}$, $i \in \{1, \ldots, n \}$, correspond to spatial units situated on either a regular or irregular lattice within a geographic domain $\mathcal{D} \subset \mathbb{R}^d$, with $d \geq 1$, the spatial scalar-on-function regression model (SSoFRM) takes the following form:
\begin{align}
\mathbb{E} \big\{ Y \mid \bm{W} Y, \X(u) \big\} &= \rho \bm{W} Y + \int_{\mathcal{I}} \X(u) \beta(u) \, du \notag\\ 
&= (\mathbb{I}_n - \rho \bm{W})^{-1} \left\lbrace \int_{\mathcal{I}} \X(u) \beta(u) \, du \right\rbrace, \label{eq2}
\end{align}
where $\mathbb{I}_n$ is an $n \times n$ diagonal matrix and $\rho \in (-1,1)$ is an unknown spatial autoregression parameter that quantifies the strength and direction of spatial dependence. The spatial connectivity among observational units is encoded through the spatial weight matrix $\bm{W} = (w_{i i^\prime})_{n \times n}$, where each element $w_{i i^\prime}$ captures the spatial proximity or interaction between locations $i$ and $i^\prime$. The spatial weight matrix is typically row-normalized, ensuring that $w_{ii^\prime} = \frac{w_{ii^\prime}}{\sum_{i^\prime=1}^{n} w_{ii^\prime}}$, which guarantees that the row sums converge to unity. Additionally, the diagonal elements of $\bm{W}$ are set to zero to exclude self-influence. Importantly, $\bm{W}$ does not need to be symmetric, and asymmetric weight matrices may be employed depending on the dataset characteristics \citep[see, e.g.,][]{Huang2021}.

In the absence of spatial dependence (i.e., $\rho = 0$), the SSoFRM simplifies to the SoFLRM given in~\eqref{eq1}. The term $\bm{W} Y$ represents the spatially lagged dependent variable, constructed using the spatial weight matrix $\bm{W}$, which encodes the spatial structure and interactions among observational units. The inclusion of spatial lags inherently introduces endogeneity, as each unit’s response both influences and is influenced by neighboring units. Consequently, standard estimation techniques that do not account for this endogeneity may lead to inconsistent estimators. This highlights the necessity of employing specialized estimation methods tailored for spatial regression frameworks.  

Several researchers, including \cite{Huang2018}, \cite{PinedaRios2019}, \cite{math2020}, and \cite{Huang2021}, have extensively studied the SSoFRM and its parameter estimation, exploring various estimation techniques. The methods proposed in these studies provide effective approaches for estimating the mean levels of the response variable based on given regressors. However, they may be limited in capturing the effects of regressors on the response variable in regions that deviate significantly from the central tendency of the response distribution. This limitation becomes particularly pronounced in heterogeneous datasets, where variability differs across regions or population subgroups. Heteroskedasticity arising from regional disparities in environmental conditions, social interactions, or socioeconomic factors is prevalent in spatial data analysis. The intensity of spatial dependence may vary across different points in the response distribution, leading to complexities that traditional models fail to accommodate effectively. For such heterogeneous structures, employing models that explicitly address these challenges can significantly enhance the understanding of underlying patterns beyond the mean, thereby improving the robustness and generalizability of the results.

Quantile regression (QR), originally introduced by \cite{koenker1978}, provides a comprehensive characterization of the conditional distribution of the response variable by assessing the influence of regressors at different quantile levels. Unlike traditional mean regression methods, QR offers a more nuanced perspective on the relationship between variables, allowing for a detailed exploration of their effects across the entire response distribution. This unique property also facilitates the straightforward construction of pointwise prediction intervals based on various quantiles, enhancing its applicability in uncertainty quantification. One of the key advantages of QR lies in its ability to accommodate heteroskedasticity without imposing the assumption of constant variance in the response variable, making it a particularly flexible modeling framework. Furthermore, QR is distribution-free because it does not require assumptions on the error term’s distribution, making it especially effective when dealing with non-Gaussian or heavy-tailed distributions. 

As a robust statistical approach, QR demonstrates resilience against outliers in the response variable, further strengthening its reliability in empirical applications \citep[see, e.g.,][for a detailed discussion]{koenker2005}. Numerous studies have extended QR to the functional data by developing functional linear quantile regression models that integrate QR within a functional framework. These models vary depending on whether the response and/or regressors are scalar or functional \citep[see, e.g.,][]{Cardot2005, Chen2012, Yao2017, Ma2019, Beyaztas2022, Mutis2024}. However, to our knowledge, no study has yet explored a functional linear quantile regression model that explicitly accounts for spatial dependence.

In this study, we introduce the spatial scalar-on-function quantile regression model (SSoFQRM) as an alternative to the SSoFRM in~\eqref{eq2}, enabling the analysis of how regressors influence different quantiles of the response distribution. Similar to SSoFRM, the proposed SSoFQRM integrates spatial dependence into the model through spatial lags of the response variable ($\bm{W} Y$). To address the endogeneity issue introduced by spatial lags and ensure robust estimation, we develop two novel estimation procedures inspired by \cite{Kim_Muller2004} and \cite{CHERNOZHUKOV2006}. Both estimation procedures employ a two-stage approach to mitigate endogeneity. The first stage, common to both methods, involves obtaining fitted values for $\bm{W} Y$ from a reduced-form model utilizing instrumental variables. The primary distinction between the two approaches emerges in the second stage, where these fitted spatial lags are incorporated to refine the estimation. 

Based on \cite{Kim_Muller2004}, the first estimation procedure directly estimates the SSoFQRM model by substituting the fitted values from the first stage in place of $\bm{W} Y$. In contrast, the second procedure, which builds upon \cite{CHERNOZHUKOV2006}, involves an additional optimization step searching over a range of values for the spatial autocorrelation parameter. This latter approach necessitates separate model estimations for each candidate value, thereby increasing computational complexity. Despite its higher computational cost, this method offers greater flexibility in capturing spatial dependencies, enhancing the robustness and accuracy of the estimated model. We derive the $\sqrt{n}$-consistency and asymptotic normality of the proposed estimators under some regularity conditions.

In the estimation procedures we propose, the functional regressor $\X(u)$ and its spatial lags $\bm{W} \X(u)$ serve as instrumental variables to construct the reduced form for $\bm{W} Y$ \citep[see, e.g.,][for details]{Kelejian1993, Kelejian1999}. However, the functional nature of the regressor, belonging to an infinite-dimensional space while being practically observed at discrete time points, introduces a significant challenge: the estimation of fitted values for $\bm{W} Y$ and, more critically, the estimation of the proposed SSoFQRM model itself constitutes an ill-posed problem. 

To address this issue, we first project the functional regressor onto a finite-dimensional space spanned by a set of orthonormal bases, using functional principal component (FPC) analysis. This dimensionality reduction approach enables us to approximate the SSoFQRM model by replacing the infinite-dimensional functional regressor with its finite set of projections, known as functional principal component scores. This transformation not only regularizes the estimation process but also enhances computational feasibility while preserving the essential variability in the functional predictor. 

The subsequent sections of this paper are organized as follows. In Section~\ref{sec2}, we present our spatial scalar-on-function quantile regression model, where its parameter estimation is described in Section~\ref{sec3}. The estimation and predictive performance of the proposed methods are assessed via an extensive set of Monte Carlo experiments, and the results are presented in Section~\ref{sec4}. In Section~\ref{sec5}, the practical applicability of the proposed methods is further presented via an empirical data analysis. Finally, Section~\ref{sec6} concludes the paper with potential future research directions.

\section{Spatial scalar-on-function quantile regression model}\label{sec2}

We consider a random sample $\{[Y_{1}, \X_{1}(u)],\dots, [Y_n, \X_n(u)]\}$, drawn from the population $(Y, \X)$ and observed at $n$ spatial locations distributed over either a regular or irregular lattice within a geographic domain $\mathcal{D} \subset \mathbb{R}^d$, with $d \geq 1$. Without loss of generality, we assume that the elements of the functional predictor $\X_i(u)$ belong to the Hilbert space $\mathcal{L}^2(\mathcal{I})$ and that both the response and the predictor are mean-zero processes, i.e., $\mathbb{E}(Y) = \mathbb{E}\{\X(u)\} = 0$.

For a given quantile level $\tau \in (0, 1)$, let $Q_\tau \{Y \mid \bm{W} Y, \X(u) \}$ represent the conditional quantile function of the response variable $Y$ given the spatial lags of the response $\bm{W} Y$ and the functional predictor $\X(u)$. We define the proposed SSoFQRM as:
\begin{align}
Q_\tau \{Y \mid \bm{W} Y, \X(u) \} &= \rho_\tau \bm{W} Y + \int_{\mathcal{I}} \X(u) \beta_\tau(u) \, du, \label{eq3} \\
&= (\mathbb{I}_n - \rho_\tau \bm{W})^{-1} \int_{\mathcal{I}} \X(u) \beta_\tau(u) \, du, \nonumber
\end{align}
where $\rho_\tau \in (-1, 1)$ is the spatial autocorrelation parameter specific to the quantile level $\tau$, and $\beta_\tau(u) \in \mathcal{L}^2(\mathcal{I})$ is the functional regression coefficient that characterizes the influence of $\X(u)$ on the $\tau$\textsuperscript{th} quantile of $Y$. The dependence of $\rho_\tau$ and $\beta_\tau(u)$ on $\tau$ enables the model to capture variations in spatial autocorrelation and functional effects across different quantiles of the response distribution, providing a more comprehensive understanding of the relationship between predictors and response beyond the mean structure. As in the SSoFRM model defined in~\eqref{eq2}, the inclusion of the spatially lagged response term $\bm{W} Y$ accounts for spatial dependence, ensuring that the model appropriately incorporates spatial autocorrelation effects in the quantile regression framework.

The proposed SSoFQRM in~\eqref{eq3} is estimated by minimizing the check-loss function \citep{koenker1978}, which penalizes the weighted absolute deviations using the function $\varphi_{\tau}(c) = c \psi_\tau (c)$ with $\psi_\tau (c) =  \{\tau- \mathbb{1} ( c < 0 )\}$, where $\mathbb{1} (\cdot)$ denotes the binary indicator function. The estimation procedure is formulated as follows:
\begin{equation}\label{eq4}
\underset{\begin{subarray}{c}
  \rho_{\tau} \in (-1,1), \; \beta_{\tau}(u) \in \mathcal{L}^{2}(\mathcal{I})  
  \end{subarray}}{\argmin}~ \sum_{i=1}^n \varphi_{\tau} \left\lbrace 
 Y_i - \rho_{\tau} \sum_{i^\prime = 1}^{n} w_{ii^\prime} Y_{i^\prime} - \int_{\mathcal{I}} \X_{i}(u) \beta_{\tau}(u) \, du
  \right\rbrace.
\end{equation}
This optimization problem estimates $\rho_{\tau}$ and $\beta_{\tau}(u)$ by minimizing the sum of weighted quantile residuals, ensuring robustness to heteroskedasticity and outliers while capturing spatial dependencies through the lagged response term $\bm{W} Y$.

\section{Parameter estimation}\label{sec3}

As discussed in Section~\ref{sec1}, including the spatially lagged response term $\bm{W} Y$ on the right-hand side of the model introduces the endogeneity issue. This arises because it may lead to $Q_{\tau}(\varepsilon) \neq Q_{\tau}(\varepsilon \mid \bm{W} Y)$, where $\varepsilon$ denotes the error term, $Q_{\tau}(\cdot)$ represents the $\tau$\textsuperscript{th} quantile, and $Q_{\tau}(\cdot \mid\bm{W} Y)$ denotes the conditional $\tau$\textsuperscript{th} quantile given $\bm{W} Y$. This inequality serves as a formal definition of endogeneity in QR models, indicating that the error distribution at a given quantile level is influenced by the spatially lagged response. Similar to other QR models, the SSoFQRM does not inherently correct for endogeneity. If endogeneity is not properly addressed, the estimators obtained from the model may become inconsistent, undermining the validity of inference and significantly reducing the reliability of predictions.

Another challenge in estimating the SSoFQRM is that the functional predictor resides in an infinite-dimensional space. A practical strategy to address this issue is to approximate the functional predictor by projecting it onto a finite-dimensional space using \textit{pre-specified} basis expansion functions, such as wavelet, radial, and $B$-spline bases, or by employing \textit{data-driven} approaches such as FPC analysis \citep[see, e.g.,][]{RamsaySilverman2006}. In this study, we adopt the FPC approach, where the basis functions are orthogonal eigenfunctions derived from the observed data. This method provides a more informative and adaptive representation of the functional predictor while preserving its inherent structural properties, thereby enhancing the interpretability and efficiency of the model estimation.

\subsection{Dimension reduction using functional principal component analysis}

Let $\mathcal{C}(u,v) = \text{Cov} \left\lbrace \X(u), \X(v) \right\rbrace$ denote the covariance function of the functional predictor $\X(u)$, satisfying $\int_{\mathcal{I}} \int_{\mathcal{I}} \mathcal{C}^2(u,v) \, du \, dv < \infty$. By Mercer's theorem, the covariance function $\mathcal{C}(u,v)$ admits the following spectral decomposition into orthogonal eigenfunctions and eigenvalues:
\begin{equation*}
\mathcal{C}(u,v) = \sum_{m \geq 1} \lambda_{m} \phi_{m} (u) \phi_{m}(v), \qquad \forall u, v \in \mathcal{I},
\end{equation*}
where $\phi_{m}(u)$ represents the orthogonal eigenfunctions, and $\lambda_{m}$ are the associated non-negative eigenvalues, satisfying $\lambda_{m} \geq \lambda_{m+1}$ for all $m \geq 1$.

According to the Karhunen-Lo\`{e}ve expansion, the functional predictor $\X_i(u)$ can be approximated using the first $M$ leading components as:
\begin{equation*}
\X_i(u) = \sum_{m=1}^M \xi_{im} \phi_{m}(u) = \bm{\xi}_{i}^ \top \bm{\phi}(u), 
\end{equation*}
where $\xi_{im} = \int_{\mathcal{I}} \X_i(u) \phi_{m}(u) \, du$ represents the FPC scores, which project $\X_i(u)$ onto the orthogonal eigenfunctions $[\phi_1(u),\dots,\phi_M(u)]$. Similarly, the regression coefficient function $\beta_{\tau}(u)$ can be expanded in terms of the eigenfunctions as follows:
\begin{equation*}
\beta_{\tau}(u) = \sum_{m=1}^M \beta^{(\tau)}_{m} \phi_{m}(u) = \bm{\beta}_{\tau}^\top \bm{\phi}(u),
\end{equation*}
where $\beta^{(\tau)}_{m}= \int_{\mathcal{I}} \beta_{\tau}(u)\phi_{m}(u) \, du$ represents the projection of the regression coefficient function onto the eigenfunction basis. In this study, we determine the number of FPCs, $M$, using the explained variance criterion. Specifically, $M$ is selected such that the first $M$ FPCs account for at least 95\% of the total variation in the data, i.e., 
\begin{equation*}
\underset{1\leq M <n}{\argmin} ~ \left\lbrace \frac{\sum_{m=1}^M \lambda_{m}}{\sum_{m=1}^{n} \lambda_{m}} \geq 95\% \right\rbrace.
\end{equation*}

Let $\bm{\Psi}$ denote the $n \times M$ matrix of FPC scores, where each row consists of $\bm{\xi}_{i}^\top = (\xi_{i1}, \ldots, \xi_{i M})^\top$. Leveraging the orthonormality condition, $\int_{\mathcal{I}} \bm{\phi}(u) \bm{\phi}^\top(u) \, du = \mathbb{I}_{M \times M}$, the SSoFQRM in~\eqref{eq3} can be reformulated as a finite-dimensional optimization problem:
\begin{align} \label{eq5}
Q_{\tau}(Y \mid \bm{W} Y, \bm{\Psi}) &= \rho_{\tau} \bm{W} Y + \int_{\mathcal{I}} \bm{\Psi} \bm{\phi}(u) \bm{\phi}^\top(u) \bm{\beta}_{\tau} \, du \nonumber \\
&= \rho_{\tau} \bm{W} Y + \bm{\Psi} \int_{\mathcal{I}} \bm{\phi}(u) \bm{\phi}^\top(u) \, du \bm{\beta}_{\tau} \nonumber \\
&= \rho_{\tau} \bm{W} Y + \bm{\Psi} \bm{\beta}_{\tau},
\end{align}
where $\bm{\beta}_{\tau}=\left\lbrace \beta^{(\tau)}_{1}, \ldots, \beta^{(\tau)}_M \right\rbrace^\top$ represents the vector of unknown regression coefficients based on the FPC scores. Consequently, the infinite-dimensional regression problem in~\eqref{eq3} is approximated as:
\begin{equation}\label{eq6}
\underset{\begin{subarray}{c}
\rho_{\tau} \in (-1,1),~ \bm{\beta}_{\tau} \in \mathbb{R}^M ~~
\end{subarray}}{\argmin}~ \sum_{i=1}^n \varphi_{\tau} \left( 
Y_i - \rho_{\tau} \sum_{i^\prime = 1}^{n} w_{ii^\prime} Y_{i^\prime} - \bm{\Psi}_{i} \bm{\beta}_{\tau} \right).
\end{equation}

This formulation indicates that the infinite-dimensional SSoFQRM can be effectively approximated within a finite-dimensional space spanned by the orthonormal FPCs. However, since the direct estimation of~\eqref{eq6} does not account for endogeneity, we focus on addressing this issue within the finite-dimensional representation of SSoFQRM. To this end, we obtain the final estimates of the regression parameters using estimation procedures adapted from \cite{Kim_Muller2004} (``KM", hereafter) and \cite{CHERNOZHUKOV2006} (``CH", hereafter). Notably, in the first stage of these procedures, the instrumental variables ($\X(u)$ and $ \bm{W} \X(u)$) used for constructing the reduced form of $\bm{W} Y$ and obtaining its fitted values are transformed into $\bm{\Psi}$ and $\bm{W} \bm{\Psi}$, ensuring consistency within the approximated SSoFQRM.

\subsection{Estimation procedure based on \texorpdfstring{\citeauthor{Kim_Muller2004}'s \citeyearpar{Kim_Muller2004}}{} method}

To begin, let $\bm{\alpha}_{\tau} = (\rho_{\tau}, \bm{\beta}_{\tau}^\top)^\top$ represent the parameter vector of the approximate form of the SSoFQRM in~\eqref{eq5}. In the first stage, we express the reduced form of $\bm{W} Y$ using instrumental variables as follows:
\begin{equation*}
Q_{\tau}(\bm{W} Y \mid \bm{\Lambda}) = \bm{\Lambda} \bm{\Pi}_{\tau},
\end{equation*}
where $\bm{\Lambda} = (\bm{\Psi}, \bm{W} \bm{\Psi}, \bm{W}^2 \bm{\Psi}, \ldots, \bm{W}^P \bm{\Psi})$ is a matrix incorporating the FPC scores $\bm{\Psi}$ along with $P$ sets of instrumental variables $\{\bm{W}^p \bm{\Psi} \}$ for $p \in \{1, \ldots, P \}$, and $\bm{\Pi}_{\tau} = \left( \Pi_1^{(\tau)}, \ldots, \Pi_P^{(\tau)} \right)^\top$ is the corresponding vector of unknown regression coefficients. Consequently, the reduced form of the approximate SSoFQRM is given by:
\begin{equation} \label{eq8_new}
Q_{\tau}(Y \mid \bm{\Lambda}) = \bm{\Lambda} \bm{\Omega}_{\tau},
\end{equation}
where 
\begin{equation*}
\bm{\Omega}_{\tau}=  \left[ \bm{\Pi}_{\tau}, \left[ \begin{matrix}
\bm{1}_{M \times M} \\ \bm{0} \end{matrix} \right] \right] \bm{\alpha}_{\tau} = H \left( \bm{\Pi}_{\tau} \right) \bm{\alpha}_{\tau} 
\end{equation*}
represents the vector of unknown parameters.

Building upon these findings, the estimator for the approximate SSoFQRM, denoted by $\widehat{\bm{\alpha}}_{\tau} = \left( \widehat{\rho}_{\tau}, \widehat{\bm{\beta}}_{\tau}^\top \right)^\top$, is obtained by solving the following minimization problem:
\begin{equation}\label{eq9_new}
\underset{\begin{subarray}{c}
\bm{\alpha}_{\tau} \in \mathbb{R}   
\end{subarray}}{\argmin}~ \sum_{i=1}^n \varphi_{\tau} \left\lbrace 
 a Y_i + (1-a) \bm{\Lambda}_{i} \widehat{\bm{\Omega}}_{\tau} - \bm{\Lambda}_{i} H \left( \widehat{ \bm{\Pi}}_{\tau} \right) \bm{\alpha}_{\tau} 
\right\rbrace,
\end{equation}
where $a \in (0,1)$ is a pre-determined positive constant that reconstructs the response variable. Inspired by \cite{Amemiya}, we incorporate $a$ as an adjustment factor to enhance efficiency.

Furthermore, the first-stage estimators, $\widehat{\bm{\Pi}}_{\tau}$ and $\widehat{\bm{\Omega}}_{\tau}$, are obtained by solving the following optimization problems:
\begin{align*}
\underset{\begin{subarray}{c}
\bm{\Pi}_{\tau} \in \mathbb{R}^{P M}   
\end{subarray}}{\argmin}~ \sum_{i=1}^n \varphi_{\tau} \left( 
\sum_{i^\prime = 1}^{n} w_{ii^\prime} Y_{i^\prime} - \bm{\Lambda}_{i} \bm{\Pi}_{\tau} \right), \qquad
&
\underset{\begin{subarray}{c}
\bm{\Omega}_{\tau} \in \mathbb{R}^{P M}   
\end{subarray}}{\argmin}~ \sum_{i=1}^n \varphi_{\tau} \left( 
Y_i - \bm{\Lambda}_{i} \bm{\Omega}_{\tau}
\right).
\end{align*}

The selection of instrumental variables is crucial for addressing endogeneity in the spatially lagged response term, $\bm{W} Y$. While increasing the order of spatial lags $(p > 1)$ can generate more instruments, higher-order lags often weaken their effectiveness. Thus, an optimal choice of $P$ is necessary to balance instrument strength and ensure consistent estimation of $\rho$ and $\bm{\Pi}_{\tau}$. In practice, $P$ should remain low, particularly when spatial autocorrelation decays rapidly. Additionally, as $n$ grows, the truncation parameter $M$ must be adjusted accordingly to capture the underlying functional relationships \citep[see, e.g.,][]{Hoshino2024}.

We summarize the KM method in Algorithm~\ref{alg:Kim_Muller}.
\begin{algorithm}[!htb]
\caption{Estimation based on \cite{Kim_Muller2004}'s methodology.}\label{alg:Kim_Muller}
\begin{algorithmic}[1]
\State For a given quantile level $\tau$, estimate the regression coefficients $\widehat{ \bm{\Pi}}_{\tau}$ by running a QR of $\bm{W} Y$ on $\bm{\Lambda}$.
\State Perform a QR of $Y$ on $\bm{\Lambda}$ for the same $\tau$ and obtain the fitted values $\bm{\Lambda} \widehat{\bm{\Omega}}_{\tau}$.
\State Reconstruct $Y$ as $Y^{*} = a Y + (1-a) \bm{\Lambda} \widehat{\bm{\Omega}}_{\tau}$, where $a \in (0,1)$ is pre-determined to enhance efficiency. Note that if $a = 1$, then $Y^{*} = Y$, potentially violating the previous step.
\State Conduct a QR of $Y^{*}$ with respect to $\bm{\Lambda} H \left(\widehat{ \bm{\Pi}}_{\tau} \right)$ for the same $\tau$ to obtain $\widehat{\bm{\alpha}}_{\tau} = \left( \widehat{\rho}_{\tau}, \widehat{\bm{\beta}}_{\tau}^\top \right)^\top$, yielding the estimated spatial autocorrelation parameter $\widehat{\rho}_{\tau}$ and estimated regression coefficients $\widehat{\bm{\beta}}_{\tau}$.
\end{algorithmic}
\end{algorithm}

Let $\widehat{\bm{\alpha}}_\tau^{(\text{KM})} = \left\lbrace \widehat{\rho}_{\tau}^{(\text{KM})}, \left(\widehat{\bm{\beta}}_{\tau}^{(\text{KM})}\right)^\top \right\rbrace^\top$ represent the estimated parameters obtained via the KM approach, i.e., the estimates derived from solving~\eqref{eq9_new}. Then, the estimate of the regression coefficient function $\beta_\tau(u)$ using the KM approach is given by:
\begin{equation}\label{eq:estKM}
\widehat{\beta}_\tau^{(\text{KM})}(u) = \left(\widehat{\bm{\beta}}_{\tau}^{(\text{KM})}\right)^\top \bm{\phi}(u).
\end{equation}

The following theorem discusses the consistency and asymptotic normality properties of the proposed estimators $\widehat{\rho}_\tau^{(\text{KM})}$ and $\widehat{\beta}_\tau^{(\text{KM})}$.
\begin{theorem}\label{th:1}
Let $\bm{\theta}_{\tau, 0} = \{\rho_\tau, \beta_\tau (u) \}$ denote the true parameters of the SSoFQEM model and let $\widehat{\bm{\theta}}_\tau^{(\text{KM})} = \{ \widehat{\rho}_\tau^{(\text{KM})}, \widehat{\beta}_\tau^{(\text{KM})}(u) \}$ denote the KM estimates of $\bm{\theta}_{\tau, 0}$. Assume the conditions $C_1$--$C_7$ given in the online supplement file hold. Then, $\widehat{\bm{\theta}}_\tau^{(\text{KM})}$ is $\sqrt{n}$-consistent and we obtain:
\begin{equation*}
\sqrt{n} \big( \widehat{\bm{\theta}}_\tau^{(\text{KM})} - \bm{\theta}_{\tau,0} \big) \xrightarrow{d} \mathcal{GP} \big( \bm{0}, \bm{\Sigma}_{\bm{\theta}^{(\text{KM})}} \big),
\end{equation*}
where $\bm{\Sigma}_{\bm{\theta}^{(\text{KM})}}$ is the covariance operator of the asymptotic Gaussian process ($\mathcal{GP}$), represented in terms of the FPC eigenfunctions and the finite-dimensional covariance matrices. The detailed representation of $\bm{\Sigma}_{\bm{\theta}^{(\text{KM})}}$ is given in the online supplement file, along with the proof of Theorem~\ref{th:1}.
\end{theorem}

\subsection{Estimation procedure based on \texorpdfstring{\citeauthor{CHERNOZHUKOV2006}'s \citeyearpar{CHERNOZHUKOV2006}}{} method}

Compared to the simpler method proposed by \cite{Kim_Muller2004}, \cite{CHERNOZHUKOV2006} introduces a more intricate approach, establishing the necessary conditions for identifying the quantiles of the response variable through a set of conditional moment restrictions in the presence of an endogenous variable. Specifically, the response variable is linked to its conditional quantile function given $\bm{W} Y$ and $\bm{\Psi}$ via the Skorohod representation:
\begin{equation*}
Y = Q \left( \bm{W} Y, \bm{\Psi}, \bm{\varepsilon}_{\bm{W} Y} \right),
\end{equation*}
where $\bm{\varepsilon}_{\bm{W} Y} \sim \text{Uniform} (0, 1)$ represents an inseparable error or rank term that captures the unobserved heterogeneity in the response variable. This representation allows for the interpretation of structural quantile effects, which quantify the differences between quantiles under varying levels of $\bm{W} Y$, as actual causal effects on individual units while maintaining the unobserved heterogeneity fixed at a given quantile level $\tau \in (0, 1)$. 

The conditional moment restrictions $C_8$ given in the online supplement file, which facilitates the estimation of the quantile process, can be expressed as:
\begin{equation}\label{eq10_new}
\text{Pr} \left\lbrace Y \leq Q \left( \bm{W} Y, \bm{\Psi}, \tau  \right) \mid \bm{\Gamma}  \right\rbrace = \text{Pr} \left\lbrace Y < Q \left( \bm{W} Y, \bm{\Psi}, \tau  \right) \mid \bm{\Gamma}  \right\rbrace = \tau.
\end{equation}
The primary implication of~\eqref{eq10_new} is that, for each $\tau$, zero corresponds to the $\tau$\textsuperscript{th} quantile of the random variable $\bm{Q}_{\tau} \left\lbrace Y- Q \left( \bm{W} Y, \bm{\Psi}, \tau  \right) \mid \bm{\Gamma} \right\rbrace$, conditional on the $n \times (M + 1)$ dimensional matrix $\bm{\Gamma} = \left\lbrace \bm{\Psi}, f^* \left( \bm{\Lambda}, e \right) \right\rbrace$, where $f^*$ is an unknown function and $e$ is a random error component. Therefore,~\eqref{eq10_new} can be interpreted as an inverse quantile regression problem, which involves identifying a function that solves the quantile regression of $Y- Q \left( \bm{W} Y, \bm{\Psi}, \tau  \right)$ on $\bm{\Gamma}$ such that zero is the optimal solution:
\begin{equation*}
0 \in 
\underset{\begin{subarray}{c}
\bm{\gamma}_{\tau} \in \mathbb{R} 
\end{subarray}}{\argmin}~ \sum_{i=1}^n \varphi_{\tau} \left\lbrace Y_i - Q_{i} \left( \bm{W} Y, \bm{\Psi}, \tau  \right) - \bm{\Gamma}_{i} \bm{\gamma}_{\tau} \right\rbrace,
\end{equation*}
where $\bm{\gamma}_{\tau}=\left( \gamma^{(\tau)}_{1}, \ldots, \gamma^{(\tau)}_{M+1} \right)^\top$ represents the vector of unknown regression coefficients associated with $\bm{\Gamma}$. 

To simplify the estimation, we consider a basic linear parameter model where 
\begin{equation*}
Q \left(\bm{W} Y, \bm{\Psi}, \tau  \right) = \rho_{\tau} \bm{W} Y + \bm{\Psi} \bm{\beta}_{\tau}, 
\end{equation*}
as specified in~\eqref{eq5}. The first stage of estimating the relevant parameters, denoted by $\bm{\alpha}_{\tau} =\left( \rho_{\tau}, \bm{\beta}_{\tau}^\top \right)^\top$, is formulated as:
\begin{equation}\label{eq11_new}
\underset{\begin{subarray}{c}
\bm{\varsigma}_{\tau} \in (-1,1), ~~ \bm{\beta}_{\tau} \in \mathbb{R}^M ~~ \end{subarray}}{\argmin}~ \sum_{i=1}^n \varphi_{\tau} \left( Y_i - \rho_{\tau} \sum_{i^\prime = 1}^{n} w_{ii^\prime} Y_{i^\prime} - \bm{\Psi}_{i} \bm{\beta}_{\tau}- \bm{\varsigma}_{\tau} \widehat{f}^*_{i} \right) \widehat{T}_i(\bm{\Lambda}_i, \tau),
\end{equation}
where $\bm{\beta}_{\tau}$ and $\bm{\varsigma}_{\tau}$ are the unknown regression coefficients for $\bm{\Psi}$ and $\widehat{f}^*$, respectively, and $\widehat{T}_i(\bm{\Lambda}_i, \tau)$ is a positive weight function. In practice, $\widehat{T}_i(\bm{\Lambda}_i, \tau)$ is set to $\widehat{T}_i = 1$ and $\widehat{f}^*$ represents the least squares projection (or QR) of $\bm{W} Y$ based on the reduced form $ f^* \left( \bm{\Lambda}, e \right)$. 

The primary focus of the first stage in~\eqref{eq11_new} is estimating the spatial autocorrelation parameter $\rho_{\tau}$ by searching over a set of candidate values, $\left\{ \rho_{l}^{(\tau)} : l=1, \ldots, L \right\}$, to bring $\bm{\varsigma}_{\tau}$ as close to zero as possible. The optimal estimate of $\rho_{\tau}$ is selected as the value that minimizes $\bm{\varsigma}_{\tau}$. Finally, the estimation of the unknown regression coefficients $\widehat{\bm{\beta}}_{\tau}$ is determined by solving the following optimization problem:
\begin{equation}\label{eq12_new}
\underset{\begin{subarray}{c}
\bm{\beta}_{\tau} \in \mathbb{R}^M 
\end{subarray}}{\argmin}~ \sum_{i=1}^n \varphi_{\tau} \left( Y_{i} - \widehat{\rho}_{\tau} \sum_{i^\prime=1}^{n} w_{ii^\prime} Y_{i^\prime} - \bm{\Psi}_{i} \bm{\beta}_{\tau} \right).
\end{equation}
All parameter estimates for the approximate SSoFQRM are obtained as $\widehat{\bm{\alpha}}_{\tau}=\left( \widehat{\rho}_{\tau}, \widehat{\bm{\beta}}_{\tau}^\top \right)^\top$.

This procedure is outlined in Algorithm \ref{alg:CHERNOZHUKOV}.
\begin{algorithm}[!htb]
\caption{Estimation based on \cite{CHERNOZHUKOV2006}'s methodology.}\label{alg:CHERNOZHUKOV}
\begin{algorithmic}[1]
\State Run an ordinary least squares regression (or QR) of $\bm{W} Y$ on $\bm{\Lambda}$ and store the fitted values as $\widehat{f}^*$.
\State Define the set of candidate values for spatial autocorrelation as $\left\{ \rho_{l}^{(\tau)} : l=1, \ldots, L\right\}$.
\State Spatially filter $Y$ as $Y^{*} = Y - \rho_{\tau} \bm{W} Y$.
\State Perform a QR of $Y^{*}$ on $\bm{\Psi}$ and $\widehat{f}^*$ for the given $\tau$ and record the estimated parameter $\widehat{\bm{\varsigma}}_{\tau}$ for the fitted value of the spatial lag $\widehat{f}^{*}$.
\State Repeat the previous step for each candidate value $\rho_{l}^{(\tau)}$ and select the value that minimizes $\widehat{\bm{\varsigma}}_{\tau}$, setting it as the estimated spatial autocorrelation parameter $\widehat{\rho}_{\tau}$.
\State Spatially filter $Y$ as $Y^{**} = Y - \widehat{\rho}_{\tau} \bm{W} Y$.
\State Conduct a QR of $Y^{**}$ on $\bm{\Psi}$ for the given $\tau$ to estimate the regression coefficients $\widehat{\bm{\beta}}_{\tau}$. Note that different levels of $\tau$ can be used at each stage.
\State Combine $\widehat{\rho}_{\tau}$ and $\widehat{\bm{\beta}}_{\tau}$ to obtain the final estimator $\widehat{\bm{\alpha}}_{\tau} = \left(\widehat{\rho}_{\tau}, \widehat{\bm{\beta}}_{\tau}^{\top} \right)^{\top}$.
\end{algorithmic}
\end{algorithm}

Let $\widehat{\bm{\alpha}}_\tau^{(\text{CH})} = \left\lbrace \widehat{\rho}_{\tau}^{(\text{CH})}, \left(\widehat{\bm{\beta}}_{\tau}^{(\text{CH})}\right)^\top \right\rbrace^\top$ denote the parameter estimates obtained via the CH method, which are solutions to equation~\eqref{eq12_new}. Under this approach, the regression coefficient function $\beta_\tau(u)$ is estimated as follows:
\begin{equation}\label{eq:estCH}
\widehat{\beta}_\tau^{(\text{CH})}(u) = \left(\widehat{\bm{\beta}}_{\tau}^{(\text{CH})}\right)^\top \bm{\phi}(u).
\end{equation}

The following theorem discusses the consistency and asymptotic normality properties of the proposed estimators $\widehat{\rho}_\tau^{(\text{CH})}$ and $\widehat{\beta}_\tau^{(\text{CH})}$.
\begin{theorem}\label{th:2}
Let $\bm{\theta}_{\tau, 0} = \{\rho_\tau, \beta_\tau (u) \}$ denote the true parameters of the SSoFQEM model and let $\widehat{\bm{\theta}}_\tau^{(\text{CH})} = \{ \widehat{\rho}_\tau^{(\text{CH})}, \widehat{\beta}_\tau^{(\text{CH})}(u) \}$ denote the CH estimates of $\bm{\theta}_{\tau, 0}$. Assume the conditions $C_1$-$C_5$ and $C_8$-$C_{10}$ given in the online supplement file hold. Then, $\widehat{\bm{\theta}}_\tau^{(\text{CH})}$ is $\sqrt{n}$-consistent and we obtain:
\begin{equation*}
\sqrt{n} \big( \widehat{\bm{\theta}}_\tau^{(\text{CH})} - \bm{\theta}_{\tau,0} \big) \xrightarrow{d} \mathcal{GP} \big( \bm{0}, \bm{\Sigma}_{\bm{\theta}^{(\text{CH})}} \big),
\end{equation*}
where $\bm{\Sigma}_{\bm{\theta}^{(\text{CH})}}$ is the covariance operator of the asymptotic Gaussian process ($\mathcal{GP}$), represented in terms of the FPC eigenfunctions and the finite-dimensional covariance matrices. The detailed representation of $\bm{\Sigma}_{\bm{\theta}^{(\text{CH})}}$ is given in the online supplement file, along with the proof of Theorem~\ref{th:2}.
\end{theorem}

\section{Monte Carlo Experiments}\label{sec4}

We conduct a comprehensive series of Monte Carlo experiments to evaluate the estimation accuracy and predictive performance of the proposed KM and CH estimators. Specifically, we benchmark these estimators against FPC and functional partial least squares (FPLS)-based SSoFRMs \citep{Huang2021}. 

To assess the robustness of the proposed methods, we introduce a second scenario in which the data generation process includes outlier contamination. In this setting, we compare the performance of the KM and CH estimators to the robust counterparts of the FPC and FPLS-based approaches, namely RFPC and RFPLS as proposed by \cite{BSM2025}. Computationally, the \Rlogo \ code of the proposed methods is documented in the \texttt{ssofqrm} package, available at \url{https://github.com/MugeMutis/ssofqrm}.

Throughout the experiments, we consider the following process to generate the trajectories of the functional predictor. Let $ \{ \X_i(u) \}$, $i \in \{1, \ldots, n\}$ and $u \in [0,1]$, be a collection of functional observations generated from a stochastic process with a specified mean function and an Ornstein-Uhlenbeck covariance structure. The data are generated according to the following model:
\begin{equation*}
\X_i(u) = \beta_x(u) + \epsilon_i^{(x)}(u), \quad u \in [0,1], \quad i = 1, \dots, n,
\end{equation*}
where $\beta_x(u)$ is the deterministic mean function and $\epsilon_i^{(x)}(u)$ is a zero-mean stochastic process with an Ornstein-Uhlenbeck (OU) covariance kernel. The mean function is defined as 
\begin{equation*}
\beta_x(u) = \cos(2\pi u) - (u - 0.5)^2. 
\end{equation*}
The error process $\epsilon_i^{(x)}(u)$ follows a zero-mean Gaussian process with an OU covariance structure, given by $\mathbb{E}[\epsilon_i^{(x)}(u)] = 0$ and  
\begin{equation*}
\text{Cov}\left[\epsilon_i^{(x)}(u), \epsilon_i^{(x)}(v)\right] = \frac{\sigma_x^2}{2 \theta_x} e^{-\theta_x |u - v|}, \qquad \forall u,v \in [0,1].
\end{equation*}
Here, $\sigma_x^2$ is the variance parameter that controls the amplitude of fluctuations, and $\theta_x > 0$ is the decay parameter that governs the temporal correlation. We consider $\sigma^2_x = 1$ and $\theta_x = 1/37$ \citep{fda.usc}. The row-normalized spatial weight matrix $\bm{W} = (w_{ii^\prime})_{n\times n}$ is constructed based on a one-dimensional regular grid. Specifically, for $i \neq i^\prime$, the weights are defined as \begin{equation*}
w_{ii^\prime} = \frac{\frac{1}{\left| i-i^\prime \right|}}{\sum_{i^\prime=1}^n \frac{1}{ \left| i-i^\prime \right|}}, 
\end{equation*}
ensuring that each row of $\bm{W}$ sums to one. For the diagonal elements, we set $w_{ii} = 0$, meaning that no spatial influence is assigned to a location. Then, the values of spatially correlated response variables are obtained as follows:
\begin{equation}\label{eq:sims}
Y = (\mathbb{I}_n - \rho \bm{W})^{-1} \left\lbrace \int_{\mathcal{I}} \X(u) \beta(u) \, du + \epsilon \right\rbrace, 
\end{equation}
where $\epsilon \sim \mathcal{N}(0,1)$. 

\begin{figure}[!htb]
\centering
\includegraphics[width=5.83cm]{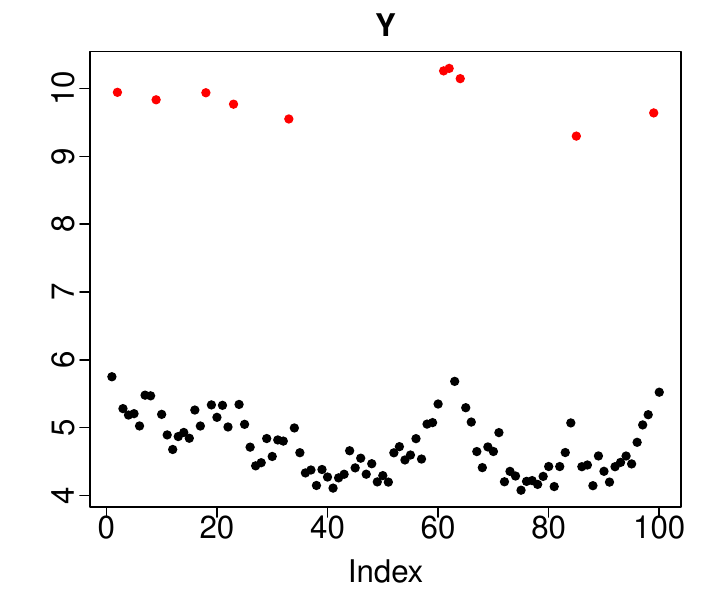}
\quad
\includegraphics[width=5.83cm]{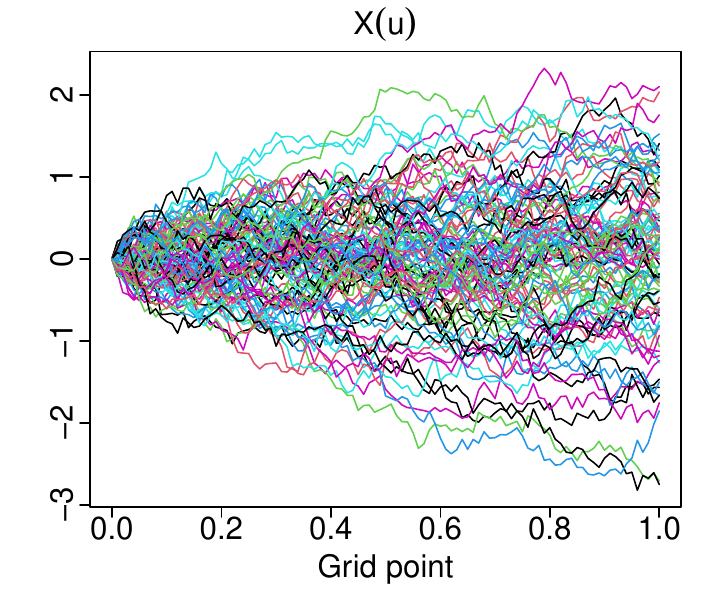}
\quad
\includegraphics[width=5.83cm]{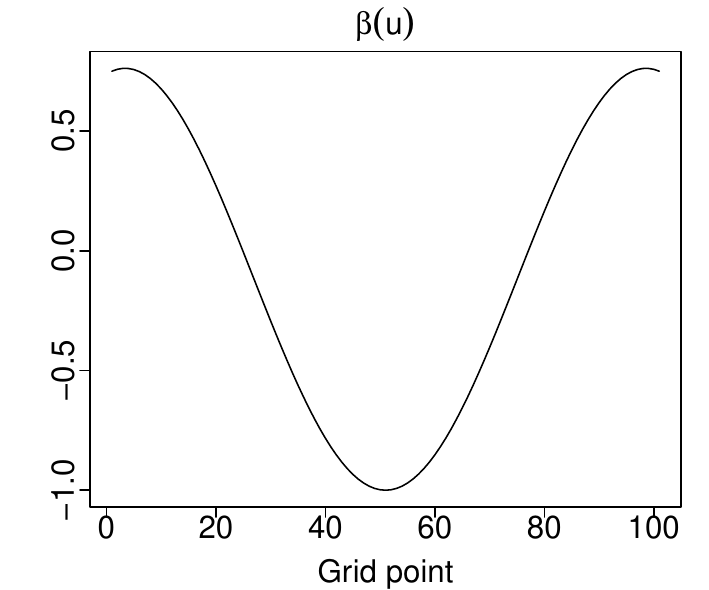}
\caption{\small{The graphs depict 100 generated sample observations for the scalar response (left panel), the functional predictor (middle panel), and the true regression coefficient function (right panel). In the left panel, black points represent normal observations, while red points indicate outliers. The data are generated under a strong spatial dependence setting with $\rho = 0.9$}.}\label{fig:Fig_1}
\end{figure}

To further assess the robustness of the proposed method, we introduce a second scenario in which the response variable is deliberately contaminated with outliers at contamination levels of 5\% and 10\%. Given that the QR captures the conditional distribution of the response variable, only the response values are affected by these outliers. Specifically, outliers are introduced by randomly selecting $n \times [5\%,~10\%]$ of the observations and replacing their corresponding error terms with values drawn from ${N}(5, 0.1)$. The contaminated response values are computed using the original model~\eqref{eq:sims}. Figure~\ref{fig:Fig_1} presents a visual depiction of the simulated dataset alongside the corresponding regression coefficient function, given a spatial dependence parameter of $\rho = 0.9$.

\newpage

To investigate the influence of spatial dependence, we consider three different values for the spatial autocorrelation parameter: $\rho \in \{0.1, 0.5, 0.9\}$, corresponding to weak, moderate, and strong spatial effects, respectively. Additionally, we assess model performance across varying training sample sizes, specifically $n \in \{50, 100, 250, 500\}$. To evaluate the generalization capability of the models, we generate an independent test set of size $n_{\text{test}} = 1000$ for each scenario. The models are first trained on the respective training datasets, and their predictive accuracy is then assessed using the test samples. 

To ensure a meaningful comparison, we set $\tau = 0.5$ in the proposed KM and CH estimators throughout the simulations, allowing us to evaluate SSoFRM against its median regression counterpart. Additionally, the proposed estimators are utilized to construct prediction intervals for the response values in the test set. Specifically, we first fit the proposed estimators to the training data at two quantile levels, $\tau_1 = 0.025$ and $\tau_2 = 0.975$. These fitted models are then applied to the test sample to generate a 95\% prediction interval for the response variable.

Across all scenarios, we conduct 500 Monte Carlo simulations to evaluate the performance of the proposed models. The models are applied to the generated datasets to estimate the model parameters and assess estimation accuracy using two key metrics: the integrated mean squared error of the estimated regression coefficient function ($\text{IMSE}(\widehat{\beta})$) and the root mean squared error of $\widehat{\rho}$ ($\text{RMSE}(\widehat{\rho})$). These metrics are defined as follows:
\begin{align*}
\text{IMSE}(\widehat{\beta}) &= \frac{1}{500} \sum_{s=1}^{500} \int_0^1 \left\lbrace \widehat{\beta}^{(s)}(u) - \beta(u) \right\rbrace^2 du,  \\
\text{RMSE}(\widehat{\rho}) &= \sqrt{\text{Var}(\widehat{\rho}) + \text{Bias}(\widehat{\rho})^2},
\end{align*}
where $\widehat{\beta}^{(s)}(u)$ denotes the estimated regression coefficient function in the $s$\textsuperscript{th} simulation. The variance and bias of $\widehat{\rho}$ are computed as:
\begin{align*}
\text{Var}(\widehat{\rho}) &= \left[\frac{1}{499} \sum_{s = 1}^{500} \left\lbrace \widehat{\rho}^{(s)} - \left(\frac{1}{500} \sum_{s = 1}^{500} \widehat{\rho}^{(s)} \right) \right\rbrace^2 \right]^{1/2}, \\
\text{Bias}(\widehat{\rho}) &= \frac{1}{500} \sum_{s = 1}^{500} \left(\widehat{\rho}^{(s)} - \rho \right),
\end{align*}
where $\widehat{\rho}^{(s)}$ represents the estimate of $\widehat{\rho}$ obtained in the $s$\textsuperscript{th} simulation. The predictive accuracy of the methods is assessed using the mean squared prediction error (MSPE), defined as:
\begin{equation*}
\text{MSPE} = \frac{1}{1000\times 500} \sum_{s=1}^{500} \sum_{i=1}^{1000} \left(\widehat{Y}^{s, \text{new}}_i - Y^{s, \text{new}}_i \right)^2,
\end{equation*}
where $Y^{s, \text{new}}_i$ denotes the $i$\textsuperscript{th} observation in the test sample during the $s$\textsuperscript{th} simulation, and $\widehat{Y}^{s, \text{new}}_i$ represents its corresponding predicted value. In our simulation studies, we ensure that the training and test samples are completely independent, meaning they do not share any common data points. This guarantees that out-of-sample predictions for a new set of functional predictors, $\X^{\text{new}}(u)$, are computed using the following formulation:
\begin{equation*}
Y^{\text{new}} = (\mathbb{I}_{n_{\text{test}}} - \widehat{\rho} \bm{W}^{\text{new}})^{-1} \int_0^1 \X^{\text{new}}(u) \widehat{\beta}(u)du,
\end{equation*}
where $\bm{W}^{\text{new}}$ is the $n_{\text{test}} \times n_{\text{test}}$ row-normalized spatial weight matrix corresponding to the test sample. However, if the training and test samples include overlapping observations, an alternative trend-corrected strategy, such as the method proposed by \cite{Goulard2017}, can be employed to adjust for potential biases in the out-of-sample predictions.

We consider two key metrics to assess the accuracy of the constructed prediction intervals: the coverage probability deviance (CPD) and the interval score. For the nominal coverage probability of 95\%, let $Q_{\tau_1}$ and $Q_{\tau_2}$ denote the estimated 2.5\% and 97.5\% quantiles of the response, respectively. These metrics are defined as follows:
\begin{align*}
\text{CPD} &= \left| 0.95 - \frac{1}{1000} \sum_{i=1}^{1000} \mathbb{1} (Q_{\tau_1} \leq Y_i \leq Q_{\tau_2}) \right|, \\
\text{score} &= \frac{1}{1000} \sum_{i=1}^{1000} \left| (Q_{\tau_2} - Q_{\tau_1}) + \frac{2}{0.05} (Q_{\tau_1} - Y_i) \mathbb{1} (Y_i < Q_{\tau_1}) + \frac{2}{0.05} (Y_i - Q_{\tau_2}) \mathbb{1} (Y_i > Q_{\tau_2}) \right|.
\end{align*}
The CPD measures the absolute deviation between the nominal and empirical coverage probabilities of the constructed prediction intervals. Lower CPD values indicate that the prediction interval successfully covers most observations in the test sample. Meanwhile, the interval score evaluates both the coverage probability and the width of the prediction interval. Lower interval score values indicate more precise and narrower prediction intervals, balancing both accuracy and efficiency in predictive uncertainty estimation. 

For all scenarios and settings, we compute the upper trimmed mean values along with the corresponding standard errors for all metrics across all methods. Here, we consider 20\% trimming proportion as suggested by \cite{Wilcox2012}. This approach ensures a fair and meaningful comparison between the robust and non-robust estimators by mitigating the influence of atypical values, providing a more reliable assessment of the methods' relative performance.

Our results are presented in Tables~\ref{tab:tab_1}-\ref{tab:tab_4}. The results in Table~\ref{tab:tab_1} under no contamination (CL~=~0\%), the KM and CH estimators produce competitive performance in terms of RMSE and IMSE, particularly for moderate ($\rho = 0.5)$ and strong ($\rho = 0.9)$  spatial dependence. In these scenarios, the CH method consistently yields lower RMSE values than the KM method, especially for smaller sample sizes, indicating its superior accuracy in estimating the spatial autocorrelation parameter. The IMSE values of the regression coefficient function $\beta(u)$ also decrease as sample size increases, confirming the consistency of both KM and CH estimators. Furthermore, both estimators maintain low MSPE values, indicating strong out-of-sample predictive capabilities.
\begin{small}
\begin{center}
\tabcolsep 0.033in
\renewcommand{\arraystretch}{1.001}
\begin{longtable}{@{}lllcccccccccccc@{}}
\caption{Computed upper trimmed mean RMSE, IMSE, and MSPE values along with their standard errors (shown in brackets) across 500 Monte Carlo replications for a spatial autocorrelation parameter of $\rho = 0.1, 0.5, 0.9$. The results are reported for four different sample sizes ($n$) and three different contamination levels (CL). Values denoted by ($***$) indicate that they are smaller than 0.01, i.e. ($<0.01$).}\label{tab:tab_1} \\
\toprule
& & \multicolumn{4}{c}{$\text{CL}=0\%$} 
& \multicolumn{4}{c}{$\text{CL}=5\%$} 
& \multicolumn{4}{c}{$\text{CL}=10\%$} \\
\cmidrule(lr){4-7} \cmidrule(lr){8-11} \cmidrule(lr){12-15}
$\rho$ & Metric & Method & $n=50$ & $100$ & $250$ & $500$ & $50$ & $100$ & $250$ & $500$ & $50$ & $100$ & $250$ & $500$ \\
\midrule
\endfirsthead
\toprule
& & & \multicolumn{4}{c}{$\text{CL}=0\%$} 
& \multicolumn{4}{c}{$\text{CL}=5\%$} 
& \multicolumn{4}{c}{$\text{CL}=10\%$} \\
\cmidrule(lr){3-6} \cmidrule(lr){7-10} \cmidrule(lr){11-14}
$\rho$ & Metric & Method & $n=50$ & $100$ & $250$ & $500$ & $50$ & $100$ & $250$ & $500$ & $50$ & $100$ & $250$ & $500$ \\
\midrule
\endhead
\midrule
\multicolumn{15}{r}{Continued on next page} \\ 
\endfoot
\endlastfoot
0.1 & RMSE & KM & 0.68 & 0.58 & 0.43 & 0.34 & 0.35 & 0.26 & 0.18 & 0.18 & 0.20 & 0.16 & 0.14 & 0.12  \\
& & CH & 0.67 & 0.57 & 0.44 & 0.34 & 0.23 & 0.20 & 0.18 & 0.19 & 0.20 & 0.17 & 0.16 & 0.14 \\ 
& & FPCA & 0.39 & 0.27 & 0.20 & 0.16 & 0.36 & 0.37 & 0.25 & 0.195 & 0.47 & 0.34 & 0.24 & 0.20 \\
& & FPLS & 0.37 & 0.25 & 0.20 & 0.16 & 0.36 & 0.36 & 0.24 & 0.19 & 0.45 & 0.33 & 0.25 & 0.20 \\
& & RFPCA & 0.40 & 0.27 & 0.20 & 0.16 & 0.23 & 0.19 & 0.30 & 0.16 & 0.07 & 0.06 & 0.05 & 0.05 \\
& & RFPLS & 0.38 & 0.26 & 0.20 & 0.16 & 0.23 & 0.17 & 0.30 & 0.16 & 0.06 & 0.05 & 0.05 & 0.05 \\
\cmidrule{2-15}
& IMSE & KM & 0.09 & 0.07 & 0.05 & 0.05 & 0.09 & 0.07 & 0.06 & 0.05 & 0.10 & 0.07 & 0.06 & 0.05 \\
& & & (0.03) & (0.01) & (0.01) & (0.01) & (0.03) & (0.01) & (0.01) & (0.01) & (0.03) & (0.02) & (0.01) & (0.01) \\
& & CH & 0.09 & 0.07 & 0.06 & 0.06 & 0.09 & 0.08 & 0.06 & 0.06 & 0.10 & 0.08 & 0.06 & 0.06  \\
& & & (0.03) & (0.02) & (0.01) & (0.01) & (0.03) & (0.02) & (0.01) & (0.01) & (0.03) & (0.02) & (0.01) & (0.01) \\ 
& & FPCA & 0.08 & 0.06 & 0.04 & 0.04 & 1.73 & 2.12 & 1.08 & 0.61 & 4.52 & 3.59 & 1.85 & 1.06 \\
& & & (0.02) & (0.01) & (0.01) & (0.01) & (1.21) & (1.30) & (0.67) & (0.36) & (3.27) & (2.28) & (1.14) & (0.65) \\
& & FPLS & 0.16 & 0.30 & 0.23 & 0.13 & 10.65 & 19.75 & 15.54 & 9.31 & 23.62 & 36.36 & 28.18 & 15.73 \\ 
& & & (0.08) & (0.13) & (0.10) & (0.05) & (5.12) & (8.29) & (5.80) & (3.08) & (11.17) & (14.35) & (10.58) & (5.33) \\
& & RFPCA & 0.11 & 0.08 & 0.05 & 0.04 & 0.12 & 0.09 & 0.05 & 0.04 & 0.15 & 0.12 & 0.07 & 0.04 \\
& & & (0.04) & (0.03) & (0.01) & (0.01) & (0.05) & (0.03) & (0.01) & (0.01) & (0.06) & (0.05) & (0.02) & (0.01) \\
& & RFPLS & 0.16 & 0.25 & 0.20 & 0.12 & 0.16 & 0.26 & 0.20 & 0.13 & 0.18 & 0.29 & 0.20 & 0.14 \\
& & & (0.07) & (0.10) & (0.08) & (0.04) & (0.07) & (0.10) & (0.08) & (0.04) & (0.09) & (0.14) & (0.10) & (0.05) \\
\cmidrule{2-15}
& MSPE & KM & 0.01 & 0.01 & 0.01 & 0.01 & 0.01 & 0.01 & 0.01 & 0.01 & 0.01 & 0.01 & 0.01 & 0.01 \\
& & & ($***$) & ($***$) & ($***$) & ($***$) & ($***$) & ($***$) & ($***$) & ($***$) & ($***$) & ($***$) & ($***$) & ($***$) \\
& & CH & 0.01 & 0.01 & 0.01 & 0.01 & 0.01 & 0.01 & 0.01 & 0.01 & 0.01 & 0.01 & 0.01 & 0.01  \\
& & & ($***$) & ($***$) & ($***$) & ($***$) & ($***$) & ($***$) & ($***$) & ($***$) & ($***$) & ($***$) & ($***$) & ($***$) \\
& & FPCA & 0.01 & 0.01 & 0.01 & 0.01 & 0.11 & 0.13 & 0.10 & 0.09 & 0.42 & 0.04 & 0.34 & 0.33 \\
& & & ($***$) & ($***$) & ($***$) & ($***$) & (0.03) & (0.02) & (0.01) & (0.01) & (0.08) & (0.05) & (0.02) & (0.01) \\
& & FPLS & 0.01 & 0.01 & 0.01 & 0.01 & 0.21 & 0.25 & 0.17 & 0.13 & 0.65 & 0.61 & 0.48 & 0.40 \\
& & & ($***$) & ($***$) & ($***$) & ($***$) & (0.05) & (0.05) & (0.02) & (0.01) & (0.13) & (0.09) & (0.04) & (0.03) \\
& & RFPCA & 0.01 & 0.01 & 0.01 & 0.01 & 0.01 & 0.01 & 0.01 & 0.01 & 0.01 & 0.01 & 0.01 & 0.01 \\
& & & ($***$) & ($***$) & ($***$) & ($***$) & ($***$) & ($***$) & (0.01) & ($***$) & ($***$) & ($***$) & ($***$) & ($***$) \\
& & RFPLS & 0.01 & 0.01 & 0.01 & 0.01 & 0.01 & 0.01 & 0.01 & 0.01 & 0.01 & 0.01 & 0.01 & 0.01 \\
& & & ($***$) & ($***$) & ($***$) & ($***$) & ($***$) & ($***$) & (0.01) & ($***$) & ($***$) & ($***$) & ($***$) & ($***$) \\
\midrule
0.5 & RMSE & KM & 0.43 & 0.39 & 0.35 & 0.28 & 0.30 & 0.24 & 0.20 & 0.18 & 0.19 & 0.17 & 0.12 & 0.12 \\
& & CH & 0.39 & 0.37 & 0.33 & 0.27 & 0.17 & 0.16 & 0.13 & 0.13 & 0.11 & 0.10 & 0.09 & 0.09 \\
& & FPCA & 0.29 & 0.23 & 0.16 & 0.12 & 0.37 & 0.29 & 0.21 & 0.17 & 0.24 & 0.21 & 0.13 & 0.16 \\
& & FPLS & 0.29 & 0.23 & 0.16 & 0.12 & 0.36 & 0.29 & 0.21 & 0.17 & 0.24 & 0.21 & 0.13 & 0.16 \\
& & RFPCA & 0.30 & 0.23 & 0.16 & 0.13 & 0.07 & 0.04 & 0.03 & 0.03 & 0.04 & 0.02 & 0.03 & 0.01 \\
& & RFPLS & 0.29 & 0.23 & 0.16 & 0.13 & 0.07 & 0.04 & 0.03 & 0.03 & 0.04 & 0.02 & 0.04 & 0.02
 \\ 
\cmidrule{2-15}
& IMSE & KM & 0.09 & 0.07 & 0.06 & 0.06 & 0.10 & 0.08 & 0.06 & 0.06 & 0.14 & 0.10 & 0.07 & 0.06 \\
& & & (0.03) & (0.02) & (0.01) & (0.01) & (0.03) & (0.02) & (0.01) & (0.01) & (0.05) & (0.04) & (0.02) & (0.01) \\
& & CH & 0.09 & 0.07 & 0.06 & 0.06 & 0.10 & 0.07 & 0.06 & 0.06 & 0.10 & 0.08 & 0.07 & 0.05 \\
& & & (0.03) & (0.02) & (0.01) & (0.01) & (0.03) & (0.02) & (0.01) & (0.01) & (0.03) & (0.02) & (0.01) & (0.01) \\
& & FPCA & 0.08 & 0.06 & 0.04 & 0.04 & 2.01 & 2.17 & 1.07 & 0.60 & 4.22 & 4.24 & 2.40 & 1.11 \\
& & & (0.02) & (0.02) & (0.01) & (0.01) & (1.34) & (1.29) & (0.68) & (0.35) & (2.76) & (2.86) & (1.58) & (0.64) \\
& & FPLS & 0.17 & 0.30 & 0.23 & 0.13 & 11.13 & 20.04 & 14.63 & 9.44 & 21.23 & 41.43 & 26.69 & 15.84 \\
& & & (0.08) & (0.13) & (0.09) & (0.04) & (5.47) & (7.94) & (5.35) & (3.31) & (9.50) & (14.93) & (11.19) & (6.96) \\
& & RFPCA & 0.11 & 0.08 & 0.05 & 0.04 & 0.12 & 0.09 & 0.05 & 0.04 & 0.15 & 0.11 & 0.07 & 0.05 \\
& & & (0.04) & (0.03) & (0.01) & (0.01) & (0.05) & (0.03) & (0.01) & (0.01) & (0.07) & (0.05) & (0.02) & (0.02) \\
& & RFPLS & 0.16 & 0.26 & 0.20 & 0.12 & 0.18 & 0.20 & 0.15 & 0.11 & 0.23 & 0.22 & 0.17 & 0.13 \\
& & & (0.07) & (0.10) & (0.08) & (0.04) & (0.08) & (0.08) & (0.05) & (0.03) & (0.11) & (0.10) & (0.08) & (0.06) \\  
\cmidrule{2-15}
& MSPE & KM & 0.15 & 0.02 & 0.01 & 0.01 & 0.05 & 0.03 & 0.02 & 0.02 & 0.05 & 0.04 & 0.02 & 0.02 \\
& & & (0.44) & (0.02) & ($***$) & ($***$) & (0.05) & (0.02) & (0.01) & (0.01) & (0.05) & (0.03) & (0.01) & (0.01) \\
& & CH & 0.02 & 0.01 & 0.01 & 0.01 & 0.02 & 0.02 & 0.02 & 0.01 & 0.02 & 0.02 & 0.02 & 0.02 \\
& & & (0.01) & ($***$) & ($***$) & ($***$) & (0.01) & (0.01) & (0.01) & ($***$) & (0.01) & (0.01) & (0.01) & (0.01) \\
& & FPCA & 0.01 & 0.01 & 0.01 & 0.01 & 0.22 & 0.30 & 0.25 & 0.27 & 1.09 & 1.08 & 1.04 & 1.02 \\
& & & ($***$) & ($***$) & ($***$) & ($***$) & (0.05) & (0.04) & (0.02) & (0.01) & (0.14) & (0.10) & (0.05) & (0.03) \\
& & FPLS & 0.01 & 0.01 & 0.01 & 0.01 & 0.32 & 0.42 & 0.32 & 0.30 & 1.25 & 1.33 & 1.15 & 1.07 \\
& & & ($***$) & ($***$) & ($***$) & ($***$) & (0.07) & (0.07) & (0.04) & (0.02) & (0.21) & (0.20) & (0.10) & (0.07) \\
& & RFPCA & 0.01 & 0.01 & 0.01 & 0.01 & 0.01 & 0.01 & 0.01 & 0.01 & 0.02 & 0.02 & 0.02 & 0.01 \\
& & & ($***$) & ($***$) & ($***$) & ($***$) & ($***$) & ($***$) & ($***$) & ($***$) & (0.01) & ($***$) & ($***$) & ($***$) \\
& & RFPLS & 0.01 & 0.01 & 0.01 & 0.01 & 0.01 & 0.01 & 0.01 & 0.01 & 0.02 & 0.02 & 0.02 & 0.02 \\
& & & ($***$) & ($***$) & ($***$) & ($***$) & ($***$) & ($***$) & ($***$) & ($***$) & (0.01) & ($***$) & ($***$) & (0.01) \\  
\midrule
0.9 & RMSE & KM & 0.10 & 0.09 & 0.09 & 0.09 & 0.09 & 0.09 & 0.09 & 0.09 & 0.08 & 0.08 & 0.08 & 0.08 \\
& & CH & 0.08 & 0.06 & 0.06 & 0.05 & 0.07 & 0.05 & 0.05 & 0.05 & 0.04 & 0.05 & 0.04 & 0.04 \\ 
& & FPCA & 0.27 & 0.19 & 0.11 & 0.08 & 0.51 & 0.44 & 0.25 & 0.20 & 0.55 & 0.40 & 0.20 & 0.15 \\ 
& & FPLS & 0.29 & 0.21 & 0.12 & 0.08 & 0.52 & 0.45 & 0.26 & 0.21 & 0.57 & 0.41 & 0.21 & 0.15 \\ 
& & RFPCA & 0.31 & 0.20 & 0.11 & 0.08 & 1.87 & 1.86 & 1.88 & 1.89 & 1.81 & 1.84 & 1.88 & 1.89 \\ 
& & RFPLS & 0.32 & 0.23 & 0.12 & 0.08 & 1.85 & 1.86 & 1.88 & 1.89 & 1.79 & 1.83 & 1.87 & 1.89 \\ 
\midrule
& IMSE & KM & 0.10 & 0.07 & 0.06 & 0.06 & 0.11 & 0.09 & 0.07 & 0.06 & 0.20 & 0.12 & 0.09 & 0.07 \\
& & & (0.03) & (0.02) & (0.01) & (0.01) & (0.04) & (0.03) & (0.02) & (0.01) & (0.11) & (0.05) & (0.02) & (0.02) \\  
& & CH & 0.10 & 0.07 & 0.06 & 0.06 & 0.09 & 0.07 & 0.06 & 0.05 & 0.10 & 0.08 & 0.06 & 0.05 \\
& & & (0.03) & (0.02) & (0.01) & (0.01) & (0.02) & (0.02) & (0.01) & (0.01) & (0.04) & (0.02) & (0.01) & (0.01) \\ 
& & FPCA & 0.08 & 0.06 & 0.04 & 0.04 & 2.01 & 2.25 & 0.85 & 0.62 & 3.75 & 4.11 & 1.78 & 0.99 \\
& & & (0.03) & (0.01) & (0.01) & (0.01) & (1.44) & (1.50) & (0.56) & (0.36) & (2.69) & (2.24) & (1.14) & (0.55) \\ 
& & FPLS & 0.17 & 0.30 & 0.21 & 0.13 & 9.93 & 19.25 & 13.24 & 9.20 & 24.11 & 35.54 & 24.83 & 15.64 \\
& & & (0.08) & (0.12) & (0.09) & (0.04) & (4.15) & (8.50) & (4.68) & (3.38) & (11.47) & (13.89) & (9.90) & (5.57) \\ 
& & RFPCA & 0.12 & 0.08 & 0.05 & 0.04 & 0.22 & 0.26 & 0.12 & 0.12 & 0.71 & 0.71 & 0.51 & 0.43 \\
& & & (0.04) & (0.03) & (0.01) & ($***$) & (0.13) & (0.16) & (0.05) & (0.06) & (0.55) & (0.51) & (0.40) & (0.24)  \\ 
& & RFPLS & 0.12 & 0.26 & 0.19 & 0.12 & 0.32 & 0.57 & 0.43 & 0.26 & 1.01 & 1.78 & 1.28 & 0.80 \\
& & & (0.07) & (0.10) & (0.07) & (0.04) & (0.15) & (0.26) & (0.20) & (0.13) & (0.69) & (1.15) & (0.79) & (0.49) \\  
\midrule
& MSPE & KM & 0.11 & 0.04 & 0.03 & 0.02 & 7.34 & 7.24 & 7.99 & 7.01 & 50.28 & 40.68 & 70.88 & 23.68 \\
& & & (0.12) & (0.04) & (0.03) & (0.01) & (7.35) & (7.99) & (10.16) & (15.49) & (78.92) & (60.90) & (105.17) & (49.31) \\ 
& & CH & 0.04 & 0.03 & 0.02 & 0.02 & 0.99 & 1.22 & 1.00 & 1.23 & 6.59 & 4.97 & 5.13 & 4.02 \\
& & & (0.03) & (0.02) & ($***$) & ($***$) & (0.87) & (1.21) & (0.98) & (1.20) & (7.30) & (5.23) & (5.96) & (4.85) \\ 
& & FPCA & 0.03 & 0.02 & 0.01 & 0.01 & 4.12 & 6.05 & 5.68 & 6.21 & 24.47 & 24.43 & 24.66 & 24.80 \\
& & & (0.02) & ($***$) & ($***$) & ($***$) & (0.68) & (0.62) & (0.43) & (0.29) & (1.95) & (1.39) & (0.87) & (0.54) \\ 
& & FPLS & 0.03 & 0.02 & 0.02 & 0.01 & 4.12 & 6.26 & 5.59 & 6.22 & 24.61 & 24.41 & 24.47 & 24.89 \\
& & & (0.02) & ($***$) & ($***$) & ($***$) & (0.68) & (0.70) & (0.52) & (0.31) & (2.18) & (1.87) & (1.19) & (0.76) \\ 
& & RFPCA & 0.03 & 0.02 & 0.01 & 0.01 & 0.31 & 0.40 & 0.27 & 0.09 & 2.42 & 1.57 & 0.97 & 0.22 \\
& & & (0.02) & ($***$) & ($***$) & ($***$) & (0.09) & (0.14) & (0.14) & (0.05) & (1.29) & (0.51) & (0.67) & (0.14) \\ 
& & RFPLS & 0.03 & 0.02 & 0.02 & 0.01 & 0.31 & 0.39 & 0.27 & 0.09 & 2.68 & 1.60 & 0.95 & 0.23 \\
& & & (0.02) & ($***$) & ($***$) & ($***$) & (0.10) & (0.15) & (0.15) & (0.06) & (1.66) & (0.60) & (0.68) & (0.15) \\  
\bottomrule
\end{longtable}
\end{center}
\end{small}

\newpage

Table~\ref{tab:tab_4} provides deeper insight into the accuracy and efficiency of prediction intervals based on 95\% coverage using CPD and score metrics. When there are no outliers (CL = 0\%), both KM and CH estimators produce small CPD values across all spatial autocorrelation levels and sample sizes, confirming that the constructed prediction intervals are reliable. The score values penalize wide or inaccurate intervals and also remain reasonably low in these settings. 

\begin{small}
\begin{center}
\tabcolsep 0.034in
\begin{longtable}{@{}llccccccccccccc@{}}
\caption{Computed upper trimmed mean CPD and score ($\times 10^3$) values for 95\% prediction intervals along with their standard errors (shown in brackets) across 500 Monte Carlo replications. The results are reported for three different spatial autocorrelation levels ($\rho$), four different sample sizes ($n$), and three different contamination levels (CL).}\label{tab:tab_4} \\
\toprule
& & & \multicolumn{4}{c}{$\text{CL}=0\%$} 
& \multicolumn{4}{c}{$\text{CL}=5\%$} 
& \multicolumn{4}{c}{$\text{CL}=10\%$} \\
\cmidrule(lr){4-7} \cmidrule(lr){8-11} \cmidrule(lr){12-15}
Metric & $\rho$ & Method & $n=50$ & $100$ & $250$ & $500$ & $50$ & $100$ & $250$ & $500$ & $50$ & $100$ & $250$ & $500$ \\
\midrule
\endfirsthead
\toprule
& & & \multicolumn{4}{c}{$\text{CL}=0\%$} 
& \multicolumn{4}{c}{$\text{CL}=5\%$} 
& \multicolumn{4}{c}{$\text{CL}=10\%$} \\
\cmidrule(lr){4-7} \cmidrule(lr){8-11} \cmidrule(lr){12-15}
Metric & $\rho$ & Method & $n=50$ & $100$ & $250$ & $500$ & $50$ & $100$ & $250$ & $500$ & $50$ & $100$ & $250$ & $500$ \\
\midrule
\endhead
\midrule
\multicolumn{15}{r}{Continued on next page} \\ 
\endfoot
\endlastfoot
CPD & 0.1 & KM & 0.07 & 0.05 & 0.04 & 0.03 & 0.26 & 0.21 & 0.10 & 0.04 & 0.24 & 0.10 & 0.04 & 0.04 \\ 
& & & (0.05) & (0.03) & (0.02) & (0.02) & (0.33) & (0.31) & (0.17) & (0.02) & (0.31) & (0.12) & (0.02) & (0.02) \\ 
& & CH & 0.08 & 0.06 & 0.04 & 0.04 & 0.07 & 0.04 & 0.04 & 0.04 & 0.06 & 0.04 & 0.04 & 0.04 \\ 
& & & (0.06) & (0.04) & (0.02) & (0.02) & (0.06) & (0.03) & (0.02) & (0.01) & (0.05) & (0.03) & (0.02) & (0.02)  \\
\cmidrule(lr){2-15}
& 0.5 & KM & 0.05 & 0.04 & 0.04 & 0.04 & 0.45 & 0.41 & 0.30 & 0.05 & 0.48 & 0.23 & 0.05 & 0.07 \\ 
& & & (0.03) & (0.02) & (0.01) & (0.02) & (0.43) & (0.43) & (0.38) & (0.02) & (0.42) & (0.32) & (0.03) & (0.08)  \\ 
& & CH & 0.05 & 0.04 & 0.04 & 0.04 & 0.05 & 0.04 & 0.04 & 0.05 & 0.05 & 0.05 & 0.05 & 0.05 \\ 
& & & (0.03) & (0.02) & (0.01) & (0.02) & (0.02) & (0.01) & (0.01) & (0.01) & (0.02) & (0.01) & (0.01) & (0.01)  \\
\cmidrule(lr){2-15}
& 0.9 & KM & 0.12 & 0.05 & 0.05 & 0.05 & 0.82 & 0.79 & 0.63 & 0.35 & 0.85 & 0.58 & 0.23 & 0.26 \\  
& & & (0.21) & (0.01) & (0.01) & (0.01) & (0.35) & (0.44) & (0.43) & (0.42) & (0.46) & (0.44) & (0.35) & (0.38) \\ 
& & CH & 0.05 & 0.05 & 0.05 & 0.05 & 0.08 & 0.11 & 0.12 & 0.12 & 0.24 & 0.21 & 0.21 & 0.19 \\ 
& & & (0.01) & (0.01) & (0.01) & (0.01) & (0.12) & (0.21) & (0.22) & (0.21) & (0.36) & (0.34) & (0.33) & (0.32)  \\  
\midrule
score & 0.1 & KM & 0.55 & 0.47 & 0.32 & 0.09 & 16.66 & 18.83 & 11.07 & 2.62 & 23.77 & 6.20 & 0.64 & 0.63 \\ 
& & & (0.60) & (0.57) & (0.49) & (0.06) & (25.75) & (26.45) & (13.38) & (4.69) & (43.78) & (8.58) & (0.10) & (0.08) \\ 
& & CH & 0.23 & 0.18 & 0.14 & 0.09 & 3.99 & 4.64 & 4.04 & 0.92 & 4.73 & 3.29 & 0.71 & 0.78 \\ 
& & & (0.15) & (0.14) & (0.12) & (0.06) & (2.54) & (3.33) & (3.80) & (0.75) & (3.48) & (3.71) & (0.24) & (0.27) \\
\cmidrule(lr){2-15}
& 0.5 & KM & 0.77 & 0.80 & 0.69 & 0.47 & 85.75 & 82.31 & 49.82 & 7.94  & 130.19 & 15.05 & 1.52 & 1.16 \\ 
& & & (0.64) & (0.68) & (0.68) & (0.60) & (116.70) & (123.59) & (89.19) & (12.03) & (206.85) & (27.86) & (0.94) & (0.35) \\ 
& & CH & 0.28 & 0.26 & 0.23 & 0.20 & 4.94 & 4.95 & 4.53 & 2.37 & 5.59 & 3.91 & 1.55 & 1.54 \\ 
& & & (0.15) & (0.16) & (0.16) & (0.15) & (2.35) & (3.26) & (3.82) & (2.57) & (3.35) & (3.61) & (0.76) & (0.70) \\
\cmidrule(lr){2-15}
& 0.9 & KM & 1.70 & 1.35 & 1.21 & 1.18 & 116.11 & 106.20 & 71.33 & 2.05 & 160.86 & 49.76 & 1.71 & 1.69 \\  
& & & (1.26) & (0.90) & (0.80) & (0.76) & (87.31) & (93.47) & (85.76) & (1.57) & (166.18) & (85.63) & (1.16) & (1.18) \\ 
& & CH & 0.44 & 0.41 & 0.41 & 0.40 & 6.15 & 6.76 & 6.46 & 6.41 & 7.77 & 7.75 & 7.37 & 7.41 \\ 
& & & (0.18) & (0.18) & (0.19) & (0.18) & (2.04) & (2.91) & (3.51) & (3.40) & (3.07) & (3.39) & (3.09) & (2.91) \\
\bottomrule
\end{longtable}
\end{center}
\end{small}

\vspace{-.3in}

A clear divergence emerges in the presence of contamination (CL = 5\% and 10\%). The KM estimator's CPD values increase significantly, particularly for smaller sample sizes and larger $\rho$, indicating under-coverage of prediction intervals. More notably, the corresponding interval score values become disproportionately large. This can be attributed to the score metric imposing steep penalties for prediction intervals that fail to contain outliers -- especially extreme ones. Since contaminated responses originate from a different distributional structure (e.g., $\mathcal{N}(5, 0.1)$), the intervals centered on non-contaminated estimates fail to capture them, resulting in extreme score values. Conversely, the CH estimator exhibits much better stability in both CPD and score metrics under contamination, further supporting its resilience and accurate predictive uncertainty quantification even in the presence of outliers.

Under contamination scenarios (CL~=~5\% and 10\%), classical estimators (FPCA and FPLS) exhibit significant performance deterioration, with inflated IMSE and MSPE values, particularly in high $\rho$ settings. This emphasizes their sensitivity to outliers. In contrast, the proposed KM and CH methods maintain relatively stable RMSE, IMSE, and MSPE values across contamination levels, highlighting their robustness. The robust baseline models (RFPCA and RFPLS) show resilience in contaminated settings, especially for small to moderate $\rho$. However, their performance deteriorates drastically under strong spatial dependence ($\rho = 0.9)$,  as evidenced by inflated IMSE and MSPE values. In contrast, the CH estimator continues outperforming others in both estimation accuracy and prediction reliability, suggesting its robustness in high-dependence and contaminated environments.

We compare the proposed methods with classical and robust alternatives in terms of computing times in Table~\ref{tab:tab_5}, based on a single Monte Carlo replication. Compared with the KM estimator, the CH estimator is computationally more demanding, with runtimes increasing with sample size (e.g., from 0.91 seconds at $n=50$ to 1.57 seconds at $n=500$) but remaining relatively stable across spatial autocorrelation levels. In contrast, the KM estimator is highly efficient, requiring under 0.15 seconds even for the largest sample size, owing to its simpler two-stage estimation. Classical methods (FPCA and FPLS) are the fastest overall, completing under 0.5 seconds even for $n=500$, making them preferable for large-scale but clean datasets. However, the robust versions (RFPCA and RFPLS) show drastically increased computing times under higher $\rho$ and $n$, reaching over 60 seconds at $n=500$ and $\rho=0.9$. This highlights the substantial computational cost of robustness, especially under strong spatial dependence and high dimensionality.
\begin{small}
\begin{center}
\tabcolsep 0.13in
\begin{longtable}{@{}lcccccccccccc@{}}
\caption{Elapsed computing times for the KM, CH, FPCA, FPLS, RFPCA, and RFPLS methods (in seconds). The computing times are recorded for four different training sample sizes ($n$) and thre different spatial autocorrelation parameter ($\rho)$.}\label{tab:tab_5} \\
\toprule
& \multicolumn{4}{c}{$\rho=0.1$} 
& \multicolumn{4}{c}{$\rho = 0.5$} 
& \multicolumn{4}{c}{$\rho = 0.9$} \\
\cmidrule(lr){2-5} \cmidrule(lr){6-9} \cmidrule(lr){10-13}
Method & $n=50$ & $100$ & $250$ & $500$ & $50$ & $100$ & $250$ & $500$ & $50$ & $100$ & $250$ & $500$ \\
\midrule
\endfirsthead
\toprule
& \multicolumn{4}{c}{$\rho=0.1$} 
& \multicolumn{4}{c}{$\rho = 0.5$} 
& \multicolumn{4}{c}{$\rho = 0.9$} \\
\cmidrule(lr){2-5} \cmidrule(lr){6-9} \cmidrule(lr){10-13}
Method & $n=50$ & $100$ & $250$ & $500$ & $50$ & $100$ & $250$ & $500$ & $50$ & $100$ & $250$ & $500$ \\
\midrule
\endhead
\midrule
\multicolumn{13}{r}{Continued on next page} \\ 
\endfoot
\endlastfoot
KM & 0.02 & 0.02 & 0.03 & 0.14 & 0.02 & 0.02 & 0.02 & 0.13 & 0.03 & 0.03 & 0.02 & 0.15  \\ 
CH & 0.91 & 0.97 & 1.15 & 1.57 & 0.91 & 0.97 & 1.13 & 1.45 & 0.87 & 0.99 & 1.08 & 1.53  \\ 
FPCA & 0.02 & 0.02 & 0.08 & 0.47 & 0.02 & 0.02 & 0.08 & 0.47 & 0.02 & 0.02 & 0.08 & 0.47 \\ 
FPLS & 0.04 & 0.04 & 0.11 & 0.56 & 0.03 & 0.07 & 0.13 & 0.51 & 0.02 & 0.02 & 0.12 & 0.53  \\ 
RFPCA & 0.05 & 0.07 & 1.32 & 6.18 & 0.10 & 1.75 & 4.34  & 44.42 & 0.14 & 0.82 & 10.95 & 60.15  \\ 
RFPLS & 0.08 & 0.16 & 1.54 & 5.47 & 0.14 & 1.91 & 5.55 & 41.00 & 0.18 & 1.15 & 11.82 & 62.17 \\
\bottomrule
\end{longtable}
\end{center}
\end{small}

\vspace{-.6in}

\section{Air quality application}\label{sec5}

To evaluate the empirical performance of our proposed model, we conducted a case study using real-world air quality data from the Lombardy region in northern Italy, an area characterized by high population density and industrial activity. The data are obtained via the \texttt{ARPALData} package \citep{arpaldata} in \Rlogo \ \citep{Team24}, which provides direct access to air quality measurements collected by the Regional Agency for the Protection of the Environment (ARPA) in Lombardy. 

\begin{figure}[!htb]
\centering
\includegraphics[height=9cm]{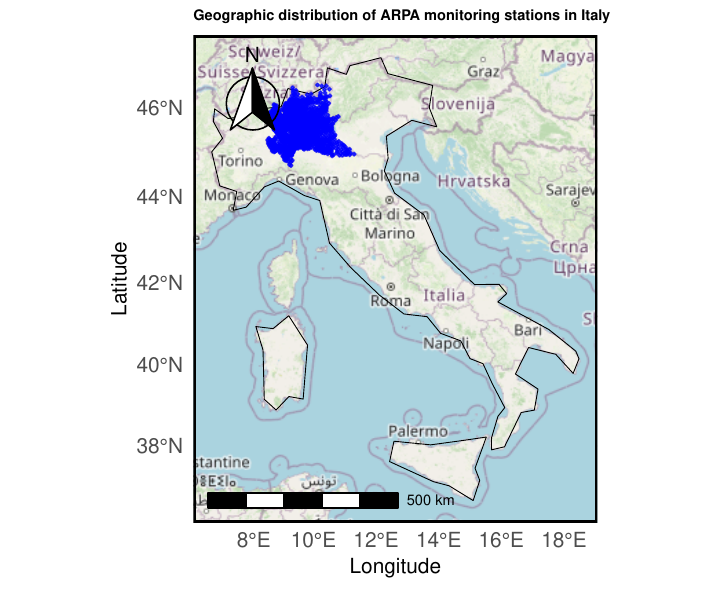}
\caption{Spatial distribution of 1481 monitoring stations (blue points) in Italy.}\label{fig:Fig_2}
\end{figure}

As shown in Figure~\ref{fig:Fig_2}, the dataset consists of observations from 1481 monitoring stations distributed across the region, with station selection based on data availability. Among the available environmental indicators, we focused on daily mean particulate matter with a diameter less than 2.5 micrometers ($\text{PM}_{2.5}$) as the response variable, given its well-established impact on public health. As the main predictor, we selected the maximum 8-hour average ozone concentration ($\text{O}3_{8-\max}$), which is a key precursor to fine particulate matter formation and a common pollutant in urban and suburban areas.

\begin{figure}[!htb]
\centering
\includegraphics[width=9cm]{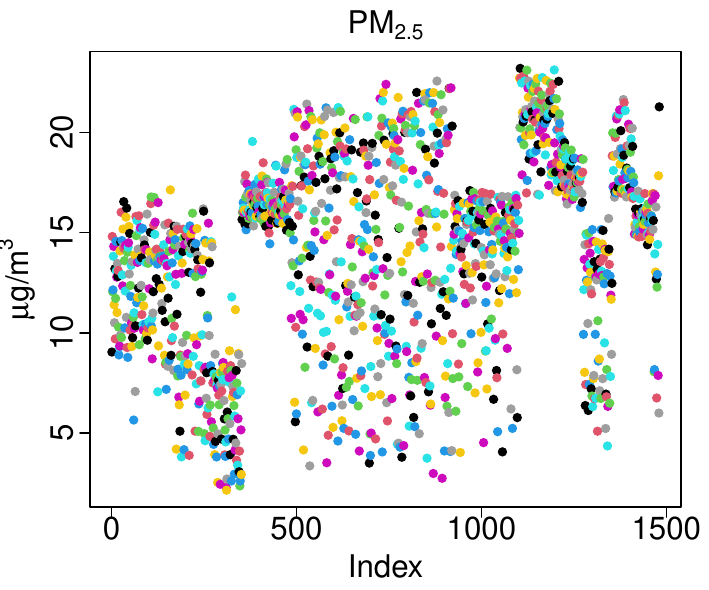}
\quad
\includegraphics[width=9cm]{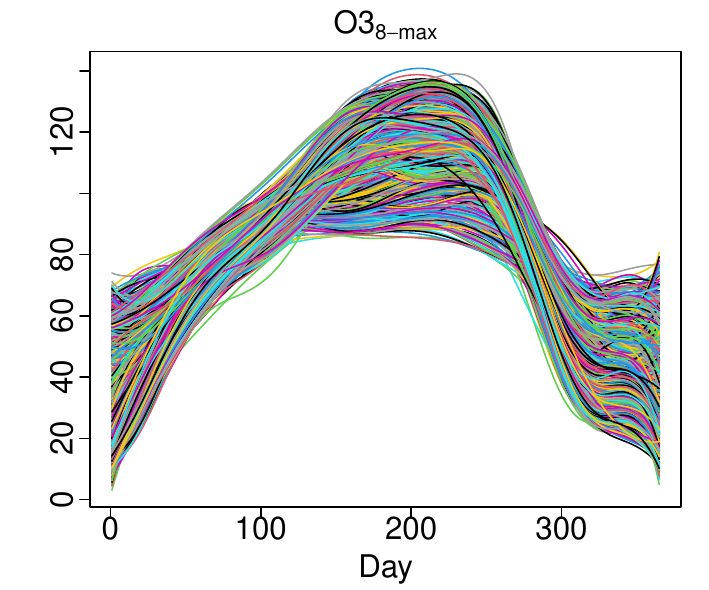}
\caption{Graphical representation of $\text{PM}_{2.5}$ (left panel) and $\text{O}3_{8-\max}$ (right panel) for 2023, based on data collected from 1481 monitoring stations in Lombardy region. Different colors indicate different monitoring stations, with observations recorded (for $\text{O}3_{8-\max}$) as functions of days $(1 \leq t,s \leq 365)$.}\label{fig:Fig_3}
\end{figure}

Our analysis spans two calendar years, with the year 2023 used as the training period and 2024 designated for out-of-sample testing. This temporal split ensures a robust assessment of model generalizability and predictive performance under real-world conditions. Modeling the relationship between $\text{PM}_{2.5}$ and $\text{O}3_{8-\max}$ is of critical importance for both environmental policy and public health planning, particularly in regions like Lombardy, where air pollution episodes are frequent. In Figure~\ref{fig:Fig_3}, we display a graphical representation of the $\text{PM}_{2.5}$ and $\text{O}3_{8-\max}$ variables collected from 1481 monitoring stations in Lombardy region; the raw $\text{O}3_{8-\max}$ data have been smoothed using 13 cubic $B$-spline basis functions for improved visualization.

Air pollution data such as $\text{PM}_{2.5}$ concentrations often exhibit strong spatial dependence due to the localized nature of emission sources, atmospheric transport mechanisms, and regional meteorological patterns. Ignoring this spatial correlation can lead to biased parameter estimates and underestimated uncertainty in prediction. Moreover, while mean regression models provide insight into average pollution levels, our proposed quantile regression may offer a more nuanced perspective by capturing the effects of predictors across the entire distribution of $\text{PM}_{2.5}$. This is particularly important for modeling upper quantiles, which correspond to extreme pollution events with the most severe health and environmental consequences.

To construct the spatial weight matrix $\bm{W} = (w_{ii^\prime})_{1481 \times 1481}$ 1481 monitoring stations in the Lombardy region, we employ a $K$-nearest neighbors approach with adaptive bi-square weights. Using each station's geographic coordinates (latitude and longitude), we compute great-circle distances $\{\text{dist}_{ii^\prime}\}$ to identify the $h$ closest neighbors of monitoring station $i$, denoted by $N^*_h(i)$, where $h$ is determined through cross-validation. The adaptive bandwidth $H^*_i$ is defined as the largest distance among these neighbors: $H^*_i = \max \{ \text{dist}_{ii^\prime} : i^\prime \in N^*_h(i) \}$. 
%\begin{equation*}
%\end{equation*}
Spatial weights are then assigned using the bi-square kernel:
\begin{equation*}
w_{ii^\prime} = \left[ 1 - \left( \frac{\text{dist}_{ii^\prime}}{H^*_i} \right)^2 \right]^2, \qquad \text{if } i^\prime \in N^{*}_{h}(i),
\end{equation*}
and $w_{ii^\prime} = 0$ otherwise. Weights are normalized so that $\sum_{i^\prime} w_{ii^\prime} = 1$ for all $i$.

\begin{figure}[!htb]
\centering
\includegraphics[width=12.5cm]{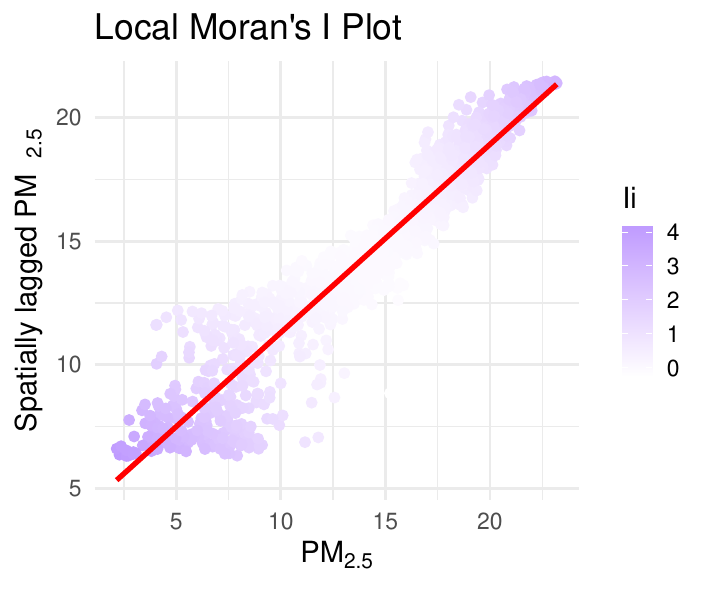}
\caption{\small{Scatter plot of Moran’s I, depicting the local spatial autocorrelation of $\text{PM}_{2.5}$ across 1481 monitoring stations in the Lombardy region}.}\label{fig:Fig_4}
\end{figure}

We compute the local Moran’s I statistic \citep{Ansellin1995} to detect spatial clustering patterns in the data. For each station $i$, the index is defined as:
\begin{equation*}
I_i = \frac{n (Y_i - \overline{Y})}{\sum_{i^{\prime} = 1}^n (Y_{i^{\prime}} - \overline{Y})^2} \sum_{i^{\prime} = 1}^n w_{i i^{\prime}} (Y_{i^{\prime}} - \overline{Y}).
\end{equation*}
This measure evaluates how a value at a specific station compares to neighboring values, indicating whether it is part of a spatial cluster. High positive values of $I_i$ reflect local similarity, while negative values suggest spatial outliers. Based on this, areas can be classified into clusters such as High-High, Low-Low, High-Low, or Low-High. Figure~\ref{fig:Fig_4} illustrates a strong positive spatial autocorrelation of $\text{PM}_{2.5}$ levels across the Lombardy region, as evidenced by the clear linear trend between observed values and their spatially lagged counterparts. The upward-sloping red regression line indicates that monitoring stations with high (or low) $\text{PM}_{2.5}$ concentrations tend to be surrounded by neighboring stations with similarly high (or low) values, confirming the presence of spatial clustering.

Leveraging the air pollution dataset, we investigate the SSoFQRM specified as:
\begin{equation*}
Q_\tau \{Y \mid \bm{W} Y, \X(u) \} = \rho_\tau \bm{W} Y + \int_1^{365} \X(u) \beta_\tau(u) \, du,
\end{equation*}
where the conditional quantile function at level $\tau$ depends on both the spatial lag of the response and a functional covariate. Parameter estimation is carried out using the proposed KM and CH estimators. For comparison, we also fit the mean-based SSoFRM using conventional techniques, including FPC, FPLS, and their robust counterparts (RFPC and RFPLS). Model predictions for $\text{PM}_{2.5}$ concentrations in 2024 are generated using observed values of $\text{O}_3^{8\text{-max}}$ from the same year. Predictive accuracy is assessed through the MSPE and the coefficient of determination ($R^2)$, defined as:
\begin{equation*}
R^2 = 1 - \frac{\sum_{i=1}^{1481} (Y_i - \widehat{Y}_i)^2}{\sum_{i=1}^{1481} (Y_i - \overline{Y})^2},
\end{equation*}
where $\overline{Y}$ denotes the sample mean of the test responses. The proposed methods are evaluated across multiple quantile levels, specifically $\tau \in \{0.5, 0.6, 0.7, 0.8, 0.9\}$, to capture distributional heterogeneity and assess model robustness in predicting upper tail pollution levels.

Our results are presented in Table~\ref{tab:tab_6}, revealing that both KM and CH estimators consistently outperform their mean-based counterparts. Across multiple quantile levels, the KM and CH methods achieve superior performance as evidenced by higher $R^2$ values and lower MSPE scores. This indicates not only strong point prediction capability but also better alignment with the observed variability in the upper tail of the pollution distribution, where the health implications are most critical. These improvements are especially pronounced at higher quantiles, affirming the model’s robustness in identifying and predicting extreme pollution episodes. 
\begin{table}[!htb]
\centering
\tabcolsep 0.072in
\caption{\small{Computed $\widehat{\rho}$, MSPE, and $R^2$ values from the air pollution data. The quantile levels used for the KM and CH methods are presented as subscripts.}}\label{tab:tab_6}
\begin{tabular}{@{}lrrrrrrrrrrrrrr@{}} 
\toprule
 & \multicolumn{5}{c}{KM} & \multicolumn{5}{c}{CH} & & & \\
\cmidrule(lr){2-6} \cmidrule(lr){7-11}
Metric & 0.5 & 0.6 & 0.7 & 0.8 & 0.9 & 0.5 & 0.6 & 0.7 & 0.8 & 0.9 & FPCA & FPLS & RFPCA & RFPLS \\
\midrule
$\widehat{\rho}$ & 0.63 & 0.57 & 0.49 & 0.52 & 0.61 & 0.62 & 0.63 & 0.61 & 0.59 & 0.66 & 0.81 & 0.48 & 0.64 & 0.70 \\
MSPE & 3.98 & 4.22 & 6.17 & 5.67 & 5.14 & 3.79 & 4.20 & 4.36 & 4.34 & 5.00 & 40.41 & 81.34 & 5.42 & 66.98 \\
$R^2$ & 0.89 & 0.90 & 0.90 & 0.88 & 0.82 & 0.89 & 0.89 & 0.89 & 0.88 & 0.81 & 0.83 & 0.73 & 0.88 & 0.80 \\
\bottomrule
\end{tabular}
\end{table}

\begin{figure}[!htb]
\centering
\includegraphics[width=5.8cm]{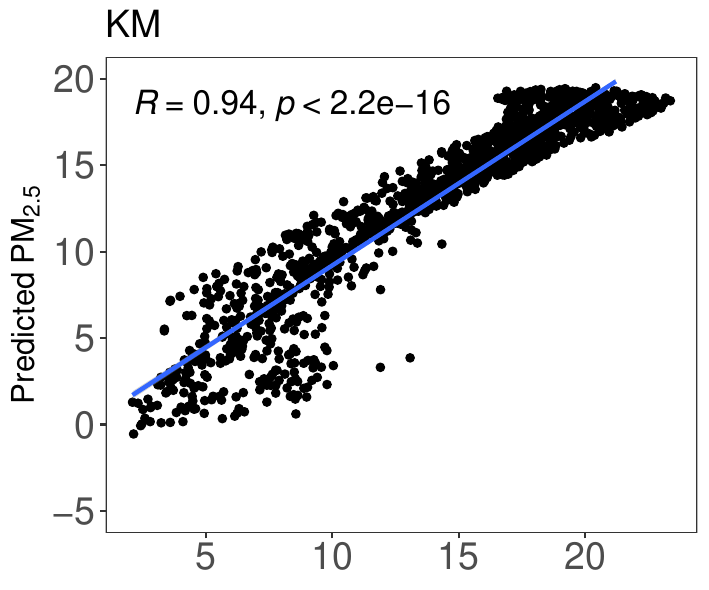}
\quad
\includegraphics[width=5.8cm]{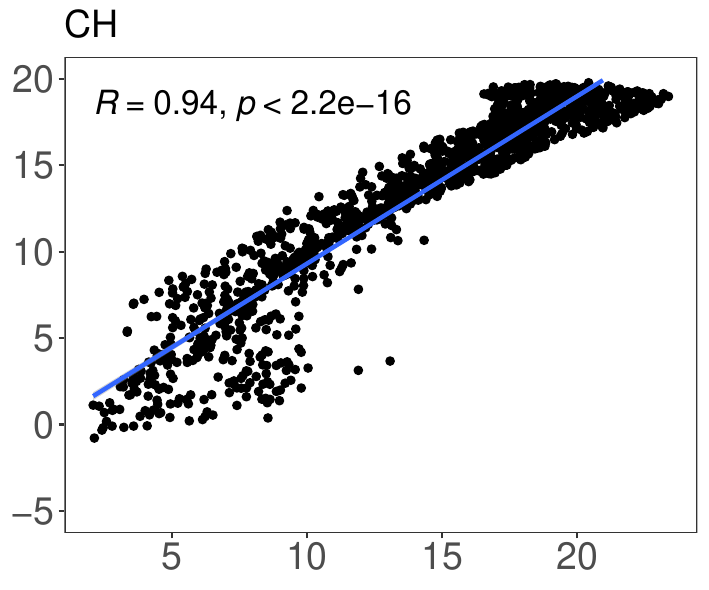}
\quad
\includegraphics[width=5.8cm]{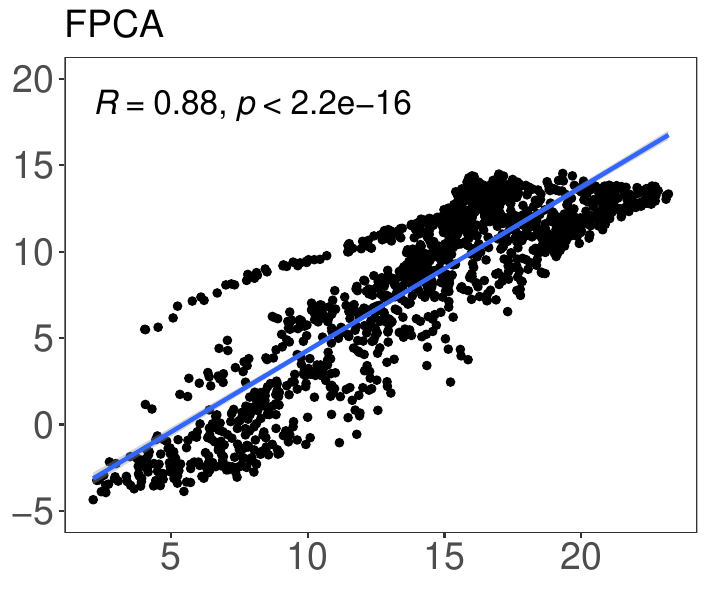}
\\  
\includegraphics[width=5.8cm]{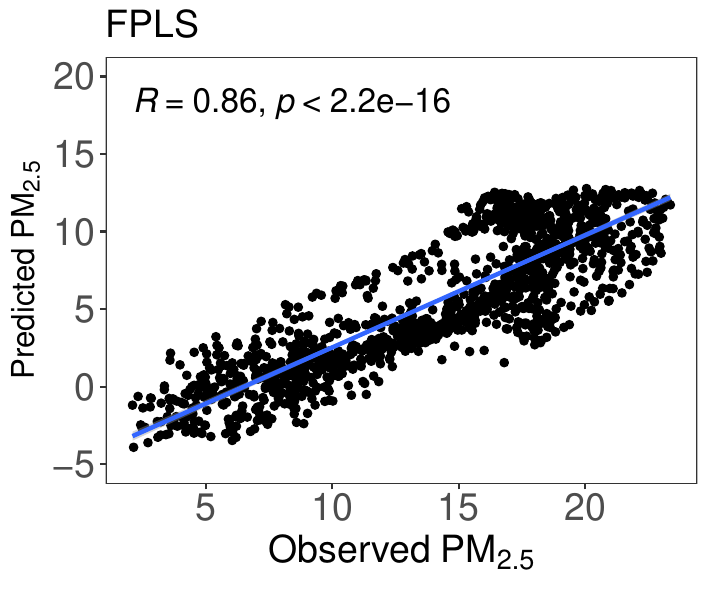}
\quad
\includegraphics[width=5.8cm]{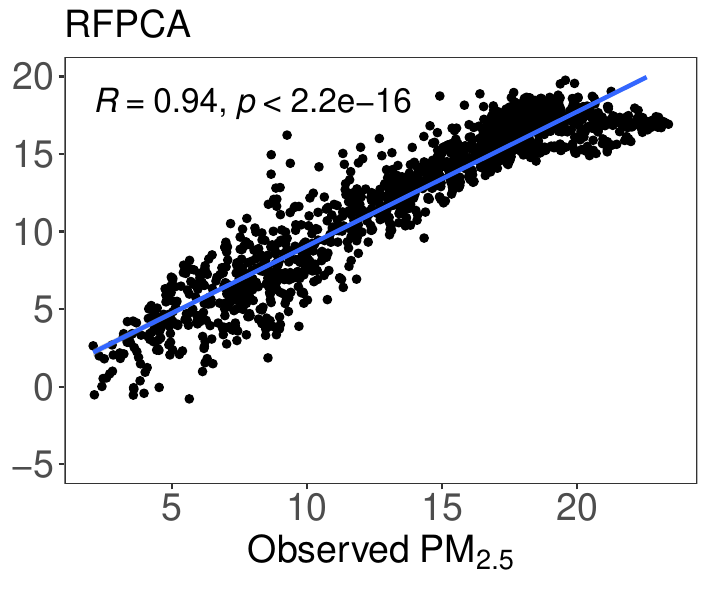}
\quad
\includegraphics[width=5.8cm]{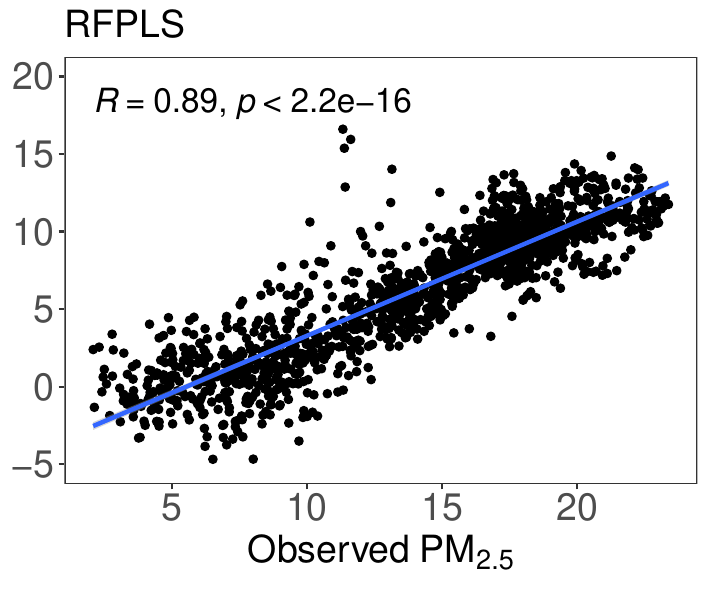}
\caption{\small{Scatter plots of observed vs. predicted $\text{PM}_{2.5}$ for the 2024 air pollution dataset. Predictions for the KM and CH are obtained using $\tau = 0.5$. Each plot includes a fitted regression line, and Pearson correlation coefficients ($R$) indicate the strength of association between observed and predicted values.}}\label{fig:Fig_5}
\end{figure}

\newpage
Figure~\ref{fig:Fig_5} complements these findings through scatterplots comparing observed and predicted $\text{PM}_{2.5}$ concentrations. The points cluster tightly around the 45-degree reference line, indicating high prediction accuracy. Notably, predictions from the quantile models maintain their accuracy even at elevated pollution levels, where mean-based models often exhibit bias or increased variance.

\begin{figure}[!htb]
\centering
\includegraphics[width=5.83cm]{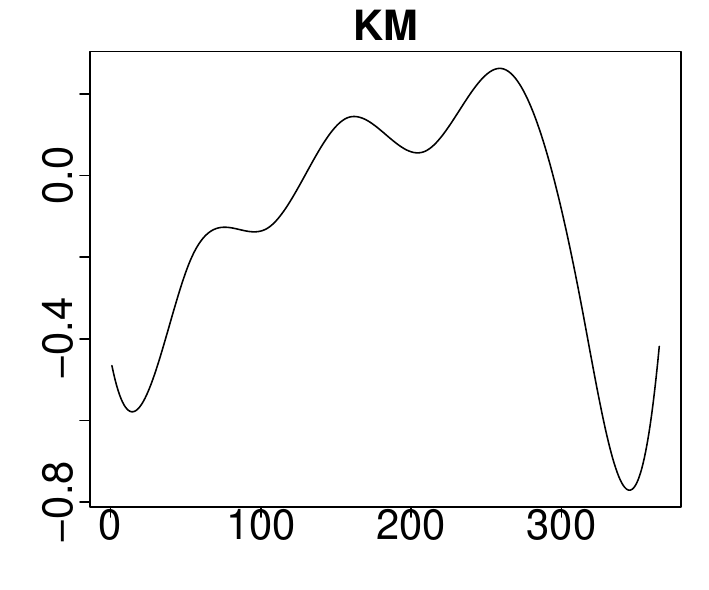}
\quad
\includegraphics[width=5.83cm]{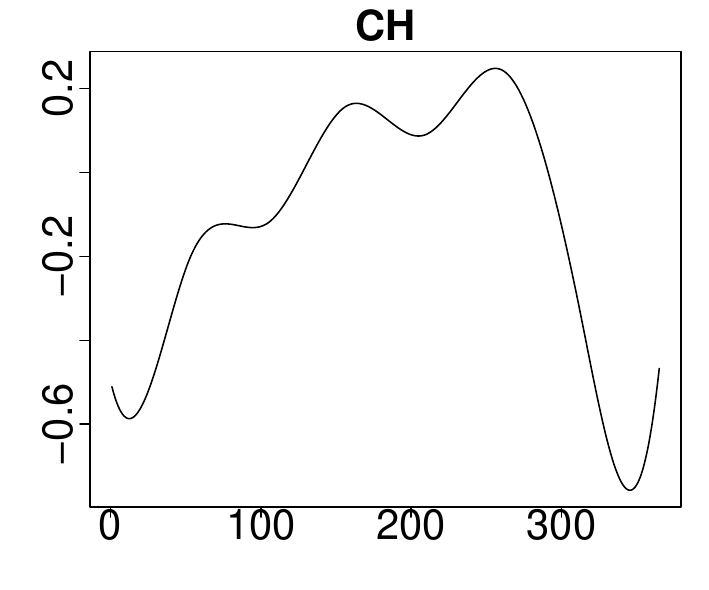}
\quad
\includegraphics[width=5.83cm]{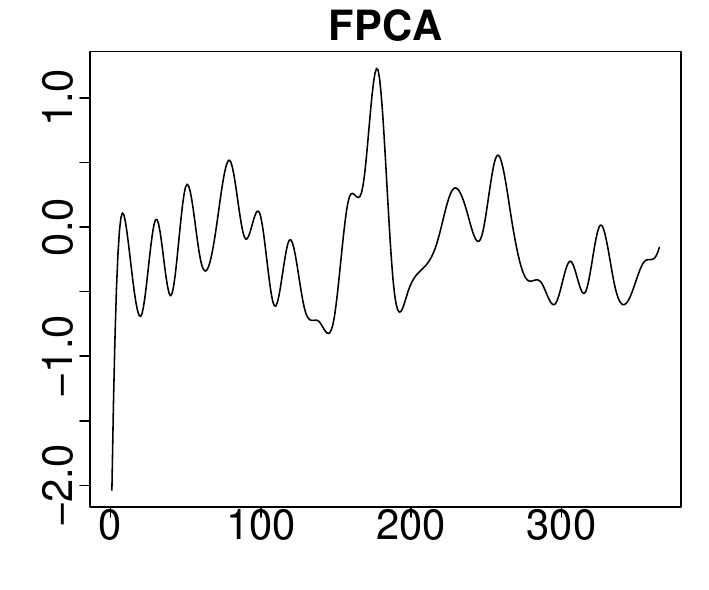}
\\
\includegraphics[width=5.83cm]{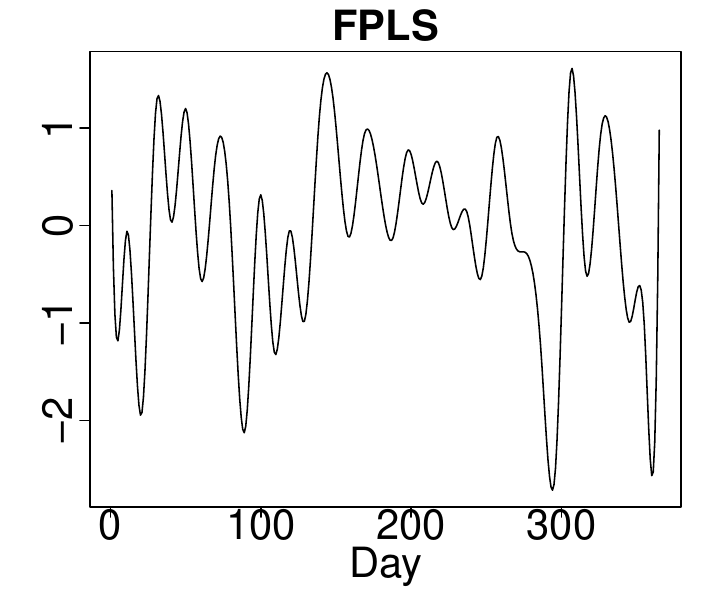}
\quad
\includegraphics[width=5.83cm]{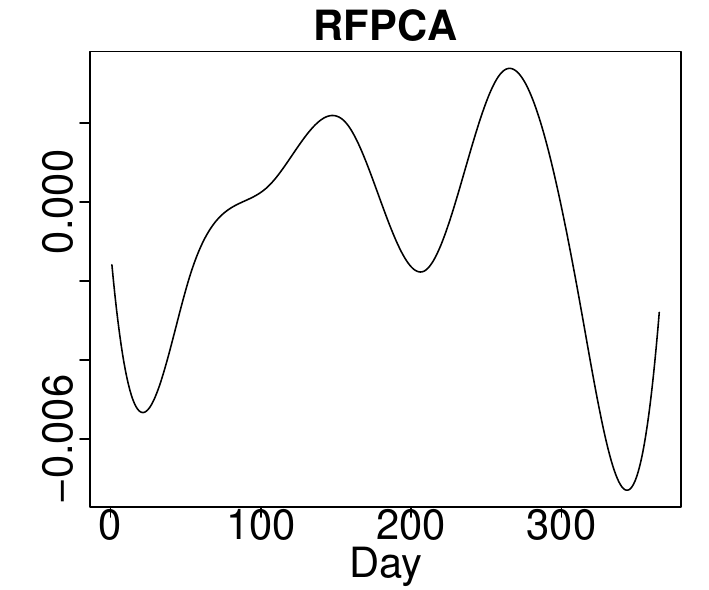}
\quad
\includegraphics[width=5.83cm]{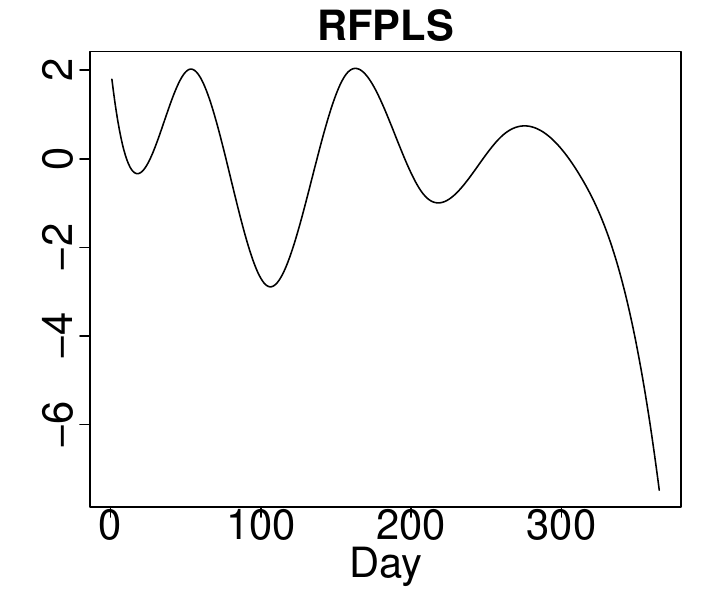}
\caption{\small{Estimated regression coefficient functions $\widehat{\beta}(u)$ for the air pollution data by all the methods. The estimated regression coefficient functions for the KM and CH are obtained using $\tau = 0.5$}.}\label{fig:Fig_6}
\end{figure}

Figure~\ref{fig:Fig_6} presents the estimated regression coefficient functions $\widehat{\beta}$(u) for all the methods. The horizontal axis in each plot represents calendar days spanning from January 1 to December 31, reflecting how the influence of ozone exposure varies over the year. The vertical axis captures the strength and direction of the effect of the daily ozone curve $\X(u)$ on the $\text{PM}_{2.5}$ response. The coefficient functions obtained from the proposed methods are non-constant, indicating that the association is not uniform throughout the year. Both methods exhibit positive coefficients during summer months (around days 150–250), aligning with the period of increased photochemical activity and ozone production. This suggests that higher ozone levels during this time significantly contribute to elevated $\text{PM}_{2.5}$ concentrations, likely due to secondary particle formation and stagnant meteorological conditions. In contrast, the coefficient functions from classical mean-based methods tend to be less interpretable or more erratic. The advantage of the proposed quantile-based approaches lies in their ability to estimate conditional effects robustly, accommodating heterogeneity in $\text{PM}_{2.5}$ responses while preserving interpretability across time.

The superior performance of the proposed quantile regression procedures can be attributed to their ability to characterize the entire conditional distribution of the response, rather than just the conditional mean. This is particularly important in environmental applications, where data often exhibit heteroskedasticity and skewness due to varying atmospheric conditions, localized sources, or measurement anomalies. By targeting different quantile levels, the KM and CH estimators offer richer insights into the underlying data structure and provide more reliable predictions for both typical and extreme conditions. These qualities make the proposed methods particularly valuable for policymakers and environmental authorities seeking to monitor and mitigate public exposure to elevated pollution levels.

\section{Conclusion}\label{sec6}

We developed a novel SSoFQRM that advances functional regression methodology by incorporating spatial dependence and allowing for the analysis of conditional quantile structures. The model is particularly well-suited for applications where the scalar response is spatially autocorrelated, and the covariates are observed as functional trajectories. To estimate the model parameters, we proposed two robust instrumental variable estimators (KM and CH), which effectively handle the endogeneity introduced by the spatially lagged response term. 

The proposed estimators' asymptotic properties, namely $\sqrt{n}$-consistency and asymptotic normality, are established under mild conditions. A series of Monte Carlo simulations confirmed their favorable performance in terms of bias, efficiency, and robustness, particularly under spatial dependence and data contamination. In the empirical application to environmental data from the Lombardy region, our methods demonstrated superior predictive accuracy in modeling $\text{PM}_{2.5}$ concentrations using ozone exposure curves. Notably, the quantile regression framework enabled more reliable predictions in the upper tails of the pollution distribution, which are of greatest concern for environmental and public health monitoring.

Future research could explore several extensions of the proposed framework. First, the model assumes that spatial dependence operates through a known spatial weight matrix, which may not fully capture complex spatial interactions or dynamic dependencies. Second, the functional predictor is treated as densely observed and noiseless, but functional data are often subject to measurement error or irregular observation. One may integrate measurement error models or regularization techniques for sparse or noisy functional data. Lastly, the proposed estimators focus on a single functional predictor, which may limit applicability in scenarios involving multiple interacting functional or scalar covariates. One direction is extending the methodology to handle multiple functional and/or scalar predictors, structured sparsity, or spatiotemporal dependencies, which would further broaden the model's scope and applicability.

\section*{Conflicts of interests/Competing interests}

The authors have no conflicts of interest to declare that are relevant to the content of this article.

\newpage
%\section*{Supplementary material}

\begin{center}
\large Supplement of ``Spatial Scalar-on-Function Quantile Regression Model"
\end{center}

In this supplement, we present the consistency and asymptotic normality of the proposed estimators. While doing so, we utilize techniques similar to those explained by \cite{Kim_Muller2004} and \cite{CHERNOZHUKOV2006}. To show the consistency and asymptotic normality of the proposed estimators, we require the following conditions.

\begin{itemize}
\item[$C_1$] The functional process $\X(u)$  admits a finite-dimensional Karhunen-Lo\`{e}ve decomposition, expressed as $\X(u) = \sum_{m=1}^M \xi_m \phi_m(u)$, where $\xi_1, \ldots, \xi_M$ are uncorrelated random variables with zero mean and variances $\lambda_m$, satisfying $\lambda_1 > \cdots > \lambda_M > 0$, and $\phi_1(u), \ldots, \phi_M(u)$ are mutually orthogonal eigenfunctions of the covariance operator, uniformly bounded over the common domain $\mathcal{I}$.
\item[$C_2$] The functional regression coefficient $\beta_\tau (u)$ resides in a finite-dimensional linear subspace of $\mathcal{L}^2(\mathcal{I})$, spanned by the orthonormal basis functions $\phi_1(u), \ldots, \phi_M(u)$, such that $\beta_{\tau}(u) = \sum_{m=1}^M \beta^{(\tau)}_{m} \phi_{m}(u)$, where the coefficients $\beta_1^{(\tau)}, \ldots, \beta_M^{(\tau)}$ determine the contribution of each basis function to $\beta_{\tau}(u)$.
\item[$C_3$] The functional process $\X$ has finite fourth moments, meaning that its norm satisfies the moment condition $\mathbb{E}(\Vert \X \Vert^4) < \mathbb{E} \left\lbrace \X^2(u) \, du \right\rbrace^2 < \infty$. 
\item[$C_4$] The matrix $\mathbb{I} - \rho_\tau \bm{W}$  is invertible for all values of $\rho_\tau$ satisfying $\vert \rho_\tau \vert < 1$.
\item[$C_5$] The absolute values of the row and column sums of the matrices $\bm{W}$ and $(\mathbb{I} - \rho_\tau \bm{W})^{-1}$ remain uniformly bounded for all $\rho_\tau$ satisfying $\vert \rho_\tau \vert < 1$.
\item[$C_6$] Let $\upsilon = Q_\tau (Y \mid \bm{\Lambda}) - \bm{\Lambda} \bm{\Omega}_\tau$ and $\zeta = Q_\tau (\bm{W} Y \mid \bm{\Lambda}) - \bm{\Lambda} \bm{\Pi}_\tau$. The sequence $\left\lbrace ( \upsilon_{i}, \zeta_{i}, \Lambda_{i} ) \right\rbrace_{i=1}^n$ consists of independent and identically distributed triplets, where $\upsilon_{i}$, $\zeta_{i}$ and $\Lambda_{i}$ denote the $i$th elements of the sequences $\upsilon$, $\zeta$ and $\bm{\Lambda}$, respectively. 
\item[$C_7$] \begin{itemize}
    \item[(i)] The third moment of $\bm{\Lambda}_i$ is finite, i.e., $\mathbb{E} \left( \Vert \bm{\Lambda}_{i}  \Vert^{3} \right) < \infty$, $\forall i$.
    \item[(ii)] The matrix $H \left( \bm{\Pi}_{\tau} \right)$ has full column rank, ensuring the identifiability of the parameter space.
    \item[(iii)] The conditional density functions $g \left(\cdot \mid \bm{\Lambda}_{i} \right)$ and $h \left( \cdot \mid \bm{\Lambda}_{i} \right)$, corresponding to $\upsilon_{i}$ and $\zeta_{i}$, are Lipschitz continuous for all $\bm{\Lambda}_{i}$. Moreover, the expectation-based matrices:, $\bm{Q}_{0 (\upsilon)} = \mathbb{E} \left\{ g\left( 0 \mid \bm{\Lambda}_{i} \right) \bm{\Lambda}_{i}^\top \bm{\Lambda}_{i} \right\}$ and $\bm{Q}_{0 (\zeta)} = \mathbb{E} \left\{ h \left( 0 \mid \bm{\Lambda}_{i} \right) \bm{\Lambda}_{i}^\top \bm{\Lambda}_{i} \right\}$ are finite and positive definite, ensuring well-posed estimation procedures.
    \item[(iv)] $\mathbb{E} \left\lbrace \psi_\tau (\upsilon_i \mid \bm{\Lambda}_i) \right\rbrace = \mathbb{E} \left\lbrace \psi_\tau (\zeta_i \mid \bm{\Lambda}_i) \right\rbrace = 0$.
\end{itemize}
\item[$C_8$] 
\begin{itemize}
    \item[(i)] For each $\bm{W} Y$, $Y = Q \left( \bm{W} Y, \bm{\Xi}, \bm{\varepsilon}_{\bm{W} Y} \right)$, where $\bm{\varepsilon}_{\bm{W} Y} \sim \text{Uniform} (0,1)$, the quantile function $Q \left( \bm{W} Y, \bm{\Xi}, \tau  \right)$ is strictly increasing in $\tau$.
    \item[(ii)] Conditional on $\bm{\Xi}$, the error term $\bm{\varepsilon}_{\bm{W} Y}$ is independent of $\bm{\Lambda} = (\bm{\Xi}, \bm{W} \bm{\Xi}, \bm{W}^2 \bm{\Xi}, \ldots, \bm{W}^P \bm{\Xi})$.
    \item[(iii)] Given the instrumental variable $\bm{\Lambda}$, the reduced form of $\bm{W} Y$ can be expressed as $\bm{W} Y = f^* \left( \bm{\Lambda}, e \right)$, where $f^*$ is an unknown function and $e$ is a random error vector.
    \item[(iv)] For each pair of $\bm{W} Y$ and $\bm{W} Y^\prime$, either $\bm{\varepsilon}_{\bm{W} Y} = \bm{\varepsilon}_{\bm{W} Y^\prime}$ or $\bm{\varepsilon}_{\bm{W} Y} \sim \bm{\varepsilon}_{\bm{W} Y^\prime}$.
\end{itemize}
\item[$C_9$] The conditional density of $Y$ is bounded such that $\sup_{Y \in \mathbb{R}} f \left( Y \mid  \bm{W} Y, \bm{\Lambda} \right ) < L_d$, where $L_d = \dim(\rho_{\tau}) + \dim(\beta_{\tau})$. Define the residual terms $r_{\text{CH}}^{(1)} = Y - \rho_{\tau} \bm{W} Y - \bm{\Xi} \bm{\beta}_{\tau} - \bm{\varsigma}_{\tau} f^* \left( \bm{\Lambda}, e \right)$, $r_{\text{CH}}^{(2)} = Y - \rho_{\tau} \bm{W} Y - \bm{\Xi} \bm{\beta}_{\tau}$, and define the followings for $ \bm{\alpha}_\tau^* \equiv \left( \rho_{\tau}, \bm{\beta}_{\tau}^\top, \bm{\varsigma}_{\tau} \right)^\top$ and  $\bm{\alpha}_{\tau} \equiv \left( \rho_{\tau}, \bm{\beta}_{\tau}^\top \right)^\top$:
\begin{align*}
\bigtriangledown \left( \bm{\alpha}_\tau^*, \tau \right) & \equiv \mathbb{E} \left\lbrace \psi_{\tau} \left( r_{\text{CH}}^{(1)} \right) \bm{\Psi}_{\tau} \right\rbrace \\
\bigtriangledown \left( \bm{\alpha}_{\tau}, \tau  \right) & \equiv \mathbb{E} \left\lbrace \psi_{\tau }\left( r_{\text{CH}}^{(2)} \right) \bm{\Psi}_{\tau} \right \}
\end{align*}
where $\bm{\Psi}_{\tau} = T (\bm{\Lambda}, \tau) \left[ f^* \left( \bm{\Lambda}_i, e_i \right), \bm{\Xi}_i^\top \right]$. Additionally, the Jacobian matrices $\frac{\partial }{\partial \left( \bm{\beta}_{\tau}^\top, \bm{\varsigma}_{\tau}  \right)} \bigtriangledown \left( \bm{\alpha}_\tau^*, \tau  \right )$ and $\frac{\partial }{\partial \left ( \rho_{\tau}, \bm{\beta}_{\tau}^\top  \right)} \bigtriangledown \left( \bm{\alpha}_{\tau}, \tau  \right )$ are continuous and have full rank uniformly over $\mathbb{U} \times \mathbb{B} \times \mathbb{C} \times (0,1)$. Moreover, the image of $\mathbb{U} \times \mathbb{B}$ under the mapping $\bm{\alpha}_{\tau} \mapsto \bigtriangledown \left( \bm{\alpha}_{\tau}  \right)$ is simply connected.
\item[$C_{10}$] With inner probability going to 1 ($\xrightarrow{p}$), the functions $\widehat{f}^*, \widehat{T} (\ell, \tau) \in \mathbb{F}$, where $\mathbb{F}$ is the class of measurable functions of $\bm{\Xi}, \bm{W} \bm{\Xi}, \bm{W}^2 \bm{\Xi}, \ldots$. Moreover, $\widehat{T} (\ell, \tau) \xrightarrow{p} T (\ell, \tau)$, $\widehat{f}^* \xrightarrow{p} f^* \left( \bm{\Lambda}, e \right )$, uniformly in $(\bm{\Lambda}, \tau)$ over compact sets, where $f^* \left( \bm{\Lambda}, e \right), T( \ell, \tau) \in \mathbb{F}$. The functions $f \left(\bm{\Lambda}, \tau \right) \in \mathbb{F}$ are uniformly smooth in $\bm{\Lambda}$ with the uniform smoothness order $ \eta > \dim \left( \bm{W} Y, \bm{\Lambda} \right )/2$, and satisfy $\left\| f \left( \bm{\Lambda}, \tau^\prime \right) - f \left( \bm{\Lambda}, \tau \right) \right\| < \mathcal{C} \left| \tau - \tau^\prime \right|^{c}$, where $\mathcal{C} > 0$ and $c>0$ for all $(\bm{\Lambda}, \tau, \tau^\prime)$.
\end{itemize}

Conditions $C_1$ and $C_2$ play a crucial role in ensuring that the infinite-dimensional SSoFQRM can be effectively approximated within a finite-dimensional space through a truncated representation involving $M$ principal components. Furthermore, the fulfillment of $C_1$ and $C_2$ is inherently linked to the validity of $C_3$, a standard assumption in the asymptotic analysis of functional principal components. This condition ensures the eigenvalues decay at an appropriate rate, allowing for a well-defined and stable finite representation. Conditions $C_4$ and $C_5$ impose structural constraints on the spatial weighting matrix, ensuring its appropriate specification within the framework of spatial regression models. These conditions are essential for maintaining the stability and identifiability of spatial dependencies in the model \citep[see, e.g.,][]{Lee2004}. Conditions $C_6$ and $C_7$ are essential for establishing the consistency and asymptotic normality of the estimators constructed in the finite-dimensional space using the KM method. Conditins $C_8$-$C_{10}$ are required for establishing consistency and asymptotic normality of the CM estimators constructed in the finite-dimensional space.

The proofs of Theorems~\ref{th:1} and~\ref{th:2} depend on the FPC decomposition of the infinite-dimensional SSoFQRM. This decomposition allows for the representation of functional predictors in a finite-dimensional space, thereby facilitating asymptotic analysis. To establish a mathematically sound framework for these proofs, we introduce a key preliminary condition that governs the FPC truncation and its associated asymptotic properties. This condition ensures that the finite-dimensional approximation retains the essential statistical characteristics of the original model, providing a valid basis for deriving consistency and asymptotic normality results.

\subsection*{Preliminary condition}

Let $\mathbb{P}$ denote the image measure induced by $\X$, defined as $\mathbb{P}(U) = P(\X \in U)$, for any Borel set $U$. Accordingly, the cumulative distribution function (CDF) of $\X$ is given by  
\begin{equation*}
\mathcal{F}(\kappa_1, \ldots, \kappa_M) := \mathbb{P}(\xi_1 \leq \kappa_1, \ldots, \xi_M \leq \kappa_M).
\end{equation*}
Next, we define the functional representation of the proposed estimator for the regression coefficient function $\widehat{\beta}_\tau(u)$ as 
\begin{equation*}
\widehat{\beta}_\tau(\mathcal{F}) = \sum_{m=1}^M \widehat{\beta}_m^{(\tau)}(\mathcal{F}) \phi_m(\mathcal{F})(u).
\end{equation*}
Under condition $C_1$, the stochastic representation of $\X(u)$ is given by  
\begin{equation*}
\X(u) = \sum_{m=1}^{M} \xi_m \phi_m(\mathcal{F})(u).
\end{equation*}
For the SSoFQRM, leveraging conditions $C_1$ and $C_2$ along with the orthonormality of the basis functions $\phi_m(\mathcal{F})(u)$, we obtain:
\begin{equation*}
Q_{\tau}(Y \mid \bm{W} Y, \bm{\Xi}) = \rho_{\tau} \bm{W} Y + \sum_{m=1}^{M} \xi_m \phi_m(\mathcal{F})(u) \phi_m^\top(\mathcal{F})(u) \beta_m^{(\tau)}.
\end{equation*}
Rewriting in matrix notation, this simplifies to:
\begin{equation*}
Q_{\tau}(Y \mid \bm{W} Y, \bm{\Xi}) = \rho_{\tau} \bm{W} Y + \bm{\Xi} \bm{\beta}_\tau(\mathcal{F}).
\end{equation*}
The results above demonstrate that, under conditions $C_1$–$C_3$, the infinite-dimensional SSoFQRM can be equivalently represented in a finite-dimensional space spanned by the FPC coefficients. Consequently, the model in \eqref{eq5} is established.

\begin{proof}[Proof of Theorem~\ref{th:1}]
Let us consider the reduced form of the approximate SSoFQRM is given in~\eqref{eq8_new}. We begin by introducing the empirical process $\mathcal{M}^(r)$, defined as
\begin{equation*} \mathcal{M}^*(r) = n^{-1/2} \sum_{i=1}^{n} \bm{\Lambda}_i \psi_\tau (a \upsilon_i - \bm{\Lambda}_i^\top r) = n^{-1/2} \sum_{i=1}^{n} m^*(\omega_i, r), 
\end{equation*}
where $r$ is $\{(P+1)M \} \times 1$ vector. The notation $\omega_i$ represents the concatenated vector $\omega_i = (\upsilon_i \bm{\Lambda}_i^\top)^\top$, and we define the function $m^*(\omega_i, r)$ as $m^*(\omega_i, r) = \bm{\Lambda}_i \psi_\tau (a \upsilon_i - \bm{\Lambda}_i^\top r)$. Furthermore, we assume that $r = n^{-1/2} \bigtriangleup$, where $\bigtriangleup \in \mathbb{R}^{(P+1)M}$, and we consider the norm $\Vert \cdot \Vert$ over this space.

Next, we define the deviation process $V^*(r)$ as $V^*(r) = n^{-1/2} \sum_{i=1}^{n} \left[ m^*(\omega_i, r) - \mathbb{E} {m^*(\omega_i, r)} \right] = \mathcal{M}^*(r) - \mathbb{E} \{\mathcal{M}^*(r) \}$. Our objective is to establish the uniform stochastic bound
\begin{equation}\label{eq:A1} 
\sup_{ \Vert r_1 - r_2 \Vert \leq N^*} \Vert V^*(r_1) - V^*(r_2) \Vert = o_p(1), 
\end{equation}
for any finite and positive constant $N^*$. The proof of~\eqref{eq:A1} relies on the following technical assumptions, adapted from \citet{Andrews1994}:
\begin{itemize} 
\item[$C_1^*$] The function $m^*(\omega_i, r)$ satisfies Pollard’s entropy condition under an envelope function $\widetilde{\mathcal{M}}(\omega_i)$. 
\item[$C_2^*$] There exists some $\delta > 2$ such that the uniform moment condition $\lim_{n \rightarrow \infty} n^{-1} \sum_{i=1}^{n} \mathbb{E} \left[\{ {\widetilde{\mathcal{M}}(\omega_i)}\}^\delta \right] < \infty$ holds asymptotically.
\end{itemize}

Define the component functions $f_1(\omega_i, r) = \bm{\Lambda}_i$, $f_2(\omega_i, r) = \psi_\tau (a \upsilon - \bm{\Lambda}_i^\top r)$, so that the function $m^*(\omega_i, r)$ can be expressed as the product $m^*(\omega_i, r) = f_1(\omega_i, r) f_2(\omega_i, r)$. Since both $f_1(\cdot, r)$ and $f_2(\cdot, r)$ belong to Type I functions with respective envelope bounds $\Vert \bm{\Lambda}_i \Vert$ and $1$ \citep[see, e.g.,][for definitions]{Andrews1994}, it follows that $m^*(\omega_i, r)$ satisfies Pollard’s entropy condition with the envelope function $\max(1, \Vert \bm{\Lambda}_i \Vert)$. This is a direct consequence of the fact that the product of two Type I functions retains the entropy condition property \citep[see, e.g., Theorems 2 and 3 in][]{Andrews1994}. Therefore, assumption $C_1^*$ is verified. Next, we establish assumption $C_2^*$. Specifically, we confirm
\begin{equation*} 
\lim_{n \rightarrow \infty} n^{-1} \sum_{i=1}^{n} \mathbb{E} \left[ \{\widetilde{\mathcal{M}}(\omega_i)\}^\delta \right] = \mathbb{E} \left[ \{\max(1, \Vert \bm{\Lambda}_i \Vert) \}^\delta \right],
\end{equation*}
which is known to be bounded by $C_7~(i)$. Consequently, the desired result in~\eqref{eq:A1} follows as a direct implication.

For any vectors $r_1, r_2 \in \mathbb{R}^{(P+1)M}$ and a positive scalar $N^* \in \mathbb{R}$, we can express them in terms of auxiliary variables $\bigtriangleup_1, \bigtriangleup_2 \in \mathbb{R}^{(P+1)M}$ and $N (>0) \in \mathbb{R}$ such that
\begin{equation*} r_1 = n^{-1/2} \bigtriangleup_1, \quad r_2 = n^{-1/2} \bigtriangleup_2, \quad \text{and} \quad N^* = n^{-1/2} N. \end{equation*}
With these notations, we define the centered process $V(\bigtriangleup) = \mathcal{M}(\bigtriangleup) - \mathbb{E} \{\mathcal{M}(\bigtriangleup)\}$ where
\begin{equation*} 
\mathcal{M}(\bigtriangleup) = n^{-1/2} \sum_{i=1}^{n} m(\omega_i, \bigtriangleup), \quad \text{with} \quad m(\omega_i, \bigtriangleup) = \bm{\Lambda}_i \psi_\tau (a \upsilon_i - n^{-1/2} \bm{\Lambda}_i^\top \bigtriangleup). 
\end{equation*}
Using this notation, the uniform stochastic bound in~\eqref{eq:A1} can be rewritten equivalently as
\begin{equation}\label{eq:A2} 
\sup_{\Vert \bigtriangleup_1 - \bigtriangleup_2 \Vert \leq N} \Vert V(\bigtriangleup_1) - V(\bigtriangleup_2) \Vert = o_p(1). \end{equation}
By choosing $\bigtriangleup_1 = \bigtriangleup$ and $\bigtriangleup_2 = 0$ in~\eqref{eq:A2}, we derive
\begin{equation}\label{eq:A3} 
\sup_{\Vert \bigtriangleup \Vert \leq N} \Vert \mathcal{M}(\bigtriangleup) - \mathcal{M}(0) - \left[ \mathbb{E} \{\mathcal{M}(\bigtriangleup)\} - \mathbb{E} \{\mathcal{M}(0)\} \right] \Vert = o_p(1). 
\end{equation}

Utilizing conditions $C_7~(i)$ and $C_7~(iii)$, it follows that $\mathbb{E} \{\mathcal{M}(\bigtriangleup)\} - \mathbb{E} \{\mathcal{M}(0)\} \rightarrow a^{-1} \bm{Q}_{0 (\upsilon)} \bigtriangleup$, as established in \citet{Kim_Muller2004}. Substituting this limiting expression into~\eqref{eq:A3} yields the refined result
\begin{equation}\label{eq:A4} 
\sup_{\Vert \bigtriangleup \Vert \leq N} \Vert \mathcal{M}(\bigtriangleup) - \mathcal{M}(0) + a^{-1} \bm{Q}_{0 (\upsilon)} \bigtriangleup \Vert = o_p(1). 
\end{equation}

We now demonstrate how the result in~\eqref{eq:A4} facilitates the introduction of the first-stage estimators. Define $\widehat{\bigtriangleup}_0 = (a-1) \sqrt{n} (\widehat{\bm{\Omega}}_\tau - \bm{\Omega}_\tau ) + \sqrt{n} (\widehat{\bm{\Pi}}_\tau - \bm{\Pi}_\tau ) \rho_\tau$. Since the terms satisfy the stochastic orders $\sqrt{n} (\widehat{\bm{\Omega}}_\tau - \bm{\Omega}_\tau) = \mathcal{O}_p(1)$ and $\sqrt{n} (\widehat{\bm{\Pi}}_\tau - \bm{\Pi}_\tau) = \mathcal{O}_p(1)$ under assumptions $C_6$ and $C_7$, it follows that $\widehat{\bigtriangleup}_0 = \mathcal{O}_p(1)$. Consequently, applying the result from~\eqref{eq:A4}, we obtain
\begin{equation*} 
\mathcal{M}(\widehat{\bigtriangleup}_0) = \mathcal{M}(0) - a^{-1} \bm{Q}_{0 (\upsilon)} \widehat{\bigtriangleup}_0 + o_p(1). 
\end{equation*}
Since for any $a > 0$, the property $\psi_\tau (a \upsilon_i) = \psi_\tau (\upsilon_i)$ holds, we express $\mathcal{M}(0) = n^{-1/2} \sum_{i=1}^{n} \bm{\Lambda}_i \psi_\tau (\upsilon_i)$. By the Lindeberg–Levy central limit theorem and given the conditions $C_6$, $C_7~(i)$, and $C_7~(iv)$, the process $\mathcal{M}(0)$ converges in distribution to a normal random variable, implying that
\begin{equation*} 
\mathcal{M}(0) = \mathcal{O}_p(1), \quad \text{and} \quad a^{-1} \bm{Q}_{0 (\upsilon)} \widehat{\bigtriangleup}_0 = \mathcal{O}_p(1). 
\end{equation*}
Thus, we conclude that 
\begin{equation}\label{eq:A5} 
\mathcal{M}(\widehat{\bigtriangleup}_0) = \mathcal{O}_p(1).
\end{equation}

Next, define the estimator
\begin{equation}\label{eq:A6} 
\widehat{\bigtriangleup}_1(\delta) = H(\widehat{\bm{\Pi}})\delta + \widehat{\bigtriangleup}_0, \quad \text{for} \quad \delta \in \mathbb{R}^{M+1}. 
\end{equation}
For some finite constant $N_1 > 0$, it follows directly from~\eqref{eq:A4} that
\begin{equation}\label{eq:A7} 
\sup_{\Vert \delta \Vert \leq N_1} \Vert \mathcal{M}\{\widehat{\bigtriangleup}_1(\delta)\} - \mathcal{M}(0) + a^{-1} \bm{Q}_{0 (\upsilon)} \widehat{\bigtriangleup}_1 (\delta) \Vert = o_p(1). 
\end{equation}
We introduce the function $\widetilde{\mathcal{M}}(\delta) = H(\widehat{\bm{\Pi}}_\tau )^\top \mathcal{M} \{\widehat{\bigtriangleup}_1(\delta) \}$, where the norm of $H(\widehat{\bm{\Pi}}\tau)$ satisfies
\begin{equation*} 
\Vert H(\widehat{\bm{\Pi}}_\tau) \Vert^2 = \text{trace} \{H(\widehat{\bm{\Pi}}_\tau) H(\widehat{\bm{\Pi}}_\tau)^\top \} = \mathcal{O}_p(1). 
\end{equation*}
Since $\widehat{\bm{\Pi}}_\tau - \bm{\Pi}_\tau = o_p(1)$, and using the arguments presented between (A.7) and (A.8) in \citet{Powel1983}, it follows that~\eqref{eq:A5} and~\eqref{eq:A7} together yield
\begin{equation}\label{eq:A8} 
\sup_{\Vert \delta \Vert \leq N_1} \Vert \widetilde{\mathcal{M}}(\delta) - H(\bm{\Pi}_\tau )^\top \mathcal{M}(\widehat{\bigtriangleup}_0) + a^{-1} \bm{Q}_z \delta \Vert = o_p(1), 
\end{equation}
where $\bm{Q}_z = H(\bm{\Pi}_\tau)^\top \bm{Q}_{0 (\upsilon)} H(\bm{\Pi}_\tau)$

Our goal now is to leverage this result to analyze the behavior of $\widehat{\delta} = \sqrt{n}(\widehat{\bm{\alpha}}_\tau^{\text{(KM)}} - \bm{\alpha}_\tau)$. To proceed, we establish that $\widehat{\delta} = \mathcal{O}_p(1)$. This follows from the application of Lemma A.4 in \citet{KoenkerZhao1996}, which holds under the following set of conditions:
\begin{itemize} 
\item[$C_3^*$] The function $\widetilde{\mathcal{M}}(\delta)$ satisfies the directional monotonicity condition $\delta^\top \widetilde{\mathcal{M}}(c^* \delta) \geq \delta^\top \widetilde{\mathcal{M}}(\delta)$ for some $c^* > 1$.
\item[$C_4^*$] The norm bound $\Vert H(\bm{\Pi}_\tau )^\top \mathcal{M} (\widehat{\bigtriangleup}_0) \Vert = \mathcal{O}_p(1)$.
\item[$C_5^*$] The convergence condition $\widetilde{\mathcal{M}}(\widehat{\delta}) = o_p(1)$, where $\widehat{\delta} = \sqrt{n}(\widehat{\bm{\alpha}}_\tau^{KM} - \bm{\alpha}_\tau)$.
\item[$C_6^*$] The matrix $\bm{Q}_z$ is positive definite.
\end{itemize}

Under the conditions established above, along with $a > 0$, Lemma A.4 in \citet{KoenkerZhao1996} provides the key results: $\widehat{\delta} = \mathcal{O}_p(1)$ and $\widehat{\delta} = a \bm{Q}_z^{-1} H(\bm{\Pi}_\tau)^\top \mathcal{M} (\widehat{\bigtriangleup}_0) + o_p(1)$. Moreover, Lemma 3.7 in \citet{KoenkerZhao1996} establishes that the function
\begin{equation*} 
h(c^*) = \sum_{i=1}^{n} \psi_\tau \{a \upsilon_i - n^{-1/2} \bm{\Lambda}_i^\top H(\widehat{\bm{\Pi}}_\tau ) \delta c^* - n^{-1/2} \bm{\Lambda}_i^\top \widehat{\bigtriangleup}_0 \} \end{equation*}
is convex. This implies that its gradient, $\delta^\top \widetilde{\mathcal{M}}(c^* \delta)$ is non-decreasing with respect to $c^*$, ensuring that condition $C_3^*$ holds.

Condition $C_4^*$ follows directly from~\eqref{eq:A5}. To verify $C_5^*$, we note that
\begin{equation*} 
\sqrt{n} \widetilde{\mathcal{M}}(\widehat{\delta}) = \left[ \left. \frac{\partial S}{\partial \bm{\alpha}_\tau} \right|_{\bm{\alpha}_\tau = \widehat{\bm{\alpha}}_\tau^{KM}} \right]_{-},
\end{equation*}
where $\left[ \left. \frac{\partial S}{\partial \bm{\alpha}_\tau} \right|_{\bm{\alpha}_\tau = \widehat{\bm{\alpha}}_\tau^{KM}} \right]_{-} = o_p(1)$ since it represents the left-hand-side partial derivatives of the objective function $S$ evaluated at the estimated solution $\widehat{\bm{\alpha}}_\tau^{KM}$. Consequently, we conclude that $\widetilde{\mathcal{M}}(\widehat{\delta}) = o_p(1)$.

Finally, condition $C_6^*$ follows straightforwardly from $C_7~(ii)$ and the identification assumption $(PM) \geq 1$. These results confirm that $\widehat{\bm{\alpha}}_\tau^{\text{(KM)}}$ is a consistent estimator of $\bm{\alpha}_\tau$, satisfying $\sqrt{n}(\widehat{\bm{\alpha}}_\tau^{\text{(KM)}} - \bm{\alpha}_\tau) = \mathcal{O}_p(1)$ with the asymptotic expansion
\begin{equation*} 
\sqrt{n}(\widehat{\bm{\alpha}}_\tau^{\text{(KM)}} - \bm{\alpha}_\tau) = a \bm{Q}_z^{-1} H(\bm{\Pi}_\tau)^\top \mathcal{M}(\widehat{\bigtriangleup}_0) + o_p(1). 
\end{equation*}

This provides the preliminary asymptotic representation of $\widehat{\bm{\alpha}}_\tau^{KM}$:
\begin{align}
\sqrt{n}(\widehat{\bm{\alpha}}_\tau^{\text{(KM)}} - \bm{\alpha}_\tau) &= \bm{Q}_z^{-1} H(\bm{\Pi}_\tau)^\top 
\Biggl\{ n^{-1/2} \sum_{i=1}^n a \bm{\Lambda}_i \psi_\tau (\upsilon_i) \nonumber \\
&\quad + (1-a) \bm{Q}_{0 (\upsilon)} n^{1/2} (\widehat{\bm{\Omega}}_\tau - \bm{\Omega}_\tau) 
- \bm{Q}_{0 (\upsilon)} n^{1/2} (\widehat{\bm{\Pi}}_\tau - \bm{\Pi}_\tau) \rho_\tau 
\Biggr\} + o_p(1). \label{eq:A9}
\end{align}

By substituting the asymptotic representations $\sqrt{n}(\widehat{\bm{\Omega}}_\tau - \bm{\Omega}_\tau) = \bm{Q}_{0 (\upsilon)} n^{-1/2} \sum_{i=1}^{n} \bm{\Lambda}_i \psi_\tau (\upsilon_i) + o_p(1)$ and $\sqrt{n} (\widehat{\bm{\Pi}}_\tau - \bm{\Pi}_\tau) = \bm{Q}_{0 (\zeta)}^{-1} n^{-1/2} \sum_{i=1}^{n} \bm{\Lambda}_i \psi_\tau (\zeta_i) + o_p(1)$ into~\eqref{eq:A9}, we observe that the first term involving $a$ cancels out with the asymptotic expansion of $\sqrt{n} (\widehat{\bm{\Omega}}_\tau - \bm{\Omega}_\tau)$ when multiplied by $-a \bm{Q}_{0 (\upsilon)}$. Simplifying further, we obtain:
\begin{align}
\sqrt{n}(\widehat{\bm{\alpha}}_\tau^{\text{(KM)}} - \bm{\alpha}_\tau) =& \ n^{-1/2} \sum_{i=1}^n \bm{Q}_z^{-1} H(\bm{\Pi}_\tau)^\top \bm{\Lambda}_i \psi_\tau (\upsilon_i) - n^{-1/2} \sum_{i=1}^n \bm{Q}_z^{-1} H(\bm{\Pi}_\tau)^\top \bm{Q}_{0 (\upsilon)} \bm{Q}_{0 (\zeta)}^{-1} \rho_\tau \bm{\Lambda}_i \psi_\tau (\zeta_i) + o_p(1), \nonumber \\
=& \ \bm{J} n^{-1/2} \sum_{i=1}^n \bm{Z}_i + o_p(1), \label{eq:A10}
\end{align}
where $\bm{J} = \bm{Q}_z^{-1} H(\bm{\Pi}_\tau)^\top \{\mathbb{1}_{PM \times PM}, \bm{Q}_{0 (\upsilon)} \bm{Q}_{0 (\zeta)}^{-1} \rho_\tau \}$ and $\bm{Z}_i = \psi_\tau (\widetilde{\mathcal{W}}_i) \otimes \bm{\Lambda}_i$ with $\psi_\tau(\widetilde{\mathcal{W}}_i) = \{\psi_\tau (\upsilon_i), \psi_\tau(\zeta_i) \}^\top$.

By $C_6$, $\bm{Z}_i$ is independently and identically distributed. In addition, from $C_7~(i)$, $\psi_\tau (\cdot)$ is bounded by 1 and nothing that $\text{Var}(\bm{Z}_i)$ is bounded; applying Lindeberg-Levy's central limit theorem leads to
\begin{equation*}
\sqrt{n} ( \widehat{\bm{\alpha}}_{\tau}^{\text{(KM)}} - \bm{\alpha}_{\tau}) \overset{d}{\rightarrow} \mathcal{N} ( 0, \bm{\Sigma}_{\bm{\alpha}^{\text{(KM)}}} ),
\end{equation*}
where $\bm{\Sigma}_{\bm{\alpha}^{\text{(CH)}}} = \bm{J} \bm{\mathcal{S}} \bm{J}^\top$ with $\bm{\mathcal{S}} = \mathbb{E} \{\psi_\tau(\widetilde{\mathcal{W}}) \bigotimes\bm{\Lambda}_i^\top \bm{\Lambda}_i\}$.

The estimate of functional regression coefficient $\widehat{\beta}_\tau^{\text{(KM)}}(u)$ is reconstructed via: $\widehat{\beta}_\tau^{\text{(KM)}}(u) = \bm{\phi}^\top(u) \widehat{\bm{\beta}}_\tau^{\text{(KM)}}$. The term $\sqrt{n} (\widehat{\rho}_\tau^{\text{(KM)}} - \rho_\tau)$ is asymptotically normal with variance $\sigma^2_{\rho^{\text{(KM)}}}$, the $(1,1)$th entry $\bm{\Sigma}_{\bm{\alpha}^{\text{(KM)}}}$. The process $\sqrt{n} \{\widehat{\beta}^{\text{(KM)}}_\tau(u) - \beta_\tau(u)\}$ converges to a Gaussian process with covariance function $\mathcal{C}^{\text{(KM)}}(u,v) = \bm{\phi}^\top(u) \bm{\Sigma}_{\bm{\beta}^{\text{(KM)}}} \bm{\phi}^\top(v)$, where $\bm{\Sigma}_{\bm{\beta}^{\text{(KM)}}}$ is the $M \times M$ submatrix of $\bm{\Sigma}_{\bm{\alpha}^{\text{(KM)}}}$ corresponding to $\bm{\beta}_\tau$. The covariance between $\sqrt{n} (\widehat{\rho}_\tau^{\text{(KM)}} - \rho_\tau)$ and $\sqrt{n} \{\widehat{\beta}^{\text{(KM)}}_\tau(u) - \beta_\tau(u)\}$ is given by:
\begin{equation*}
\text{Cov} \left[ \sqrt{n} (\widehat{\rho}_\tau^{\text{(KM)}} - \rho_\tau), \sqrt{n} \{\widehat{\beta}^{\text{(KM)}}_\tau(u) - \beta_\tau(u)\} \right] = \bm{\Sigma}_{\rho^{\text{(KM)}} \bm{\beta}^{\text{(KM)}}} \bm{\phi}(u),
\end{equation*}
where $\bm{\Sigma}_{\rho^{\text{(KM)}} \bm{\beta}^{\text{(KM)}}}$ is the $1 \times M$ block of $\bm{\Sigma}_{\bm{\alpha}^{\text{(KM)}}}$ representing the covariance between $\rho_\tau$ and $\bm{\beta}_\tau$. 

In conclusion, by conditions $C_1$-$C_7$, the estimators $\widehat{\rho}_\tau^{\text{(KM)}}$ and $\widehat{\beta}_\tau^{\text{(KM)}}(u)$ are jointly asymptotically Gaussian with:
\begin{equation*}
\sqrt{n} \begin{bmatrix} \widehat{\rho}_\tau^{\text{(KM)}} - \rho_\tau \\ \widehat{\beta}_\tau^{\text{(KM)}}(u) - \beta_\tau(u) \end{bmatrix}
\xrightarrow{d} \mathcal{GP} \left( 0, \Sigma_{\bm{\theta}^{\text{(KM)}}} \right),   
\end{equation*}
where the covariance operator $\Sigma_{\bm{\theta}^{\text{(KM)}}}$ acts on functions $f^{\text{(KM)}}(u) = \{f_{\rho}^{\text{(KM)}}, f_{\beta}^{\text{(KM)}}(u) \}$ as:
\begin{equation*}
\Sigma_{\bm{\theta}^{\text{(KM)}}} f^{\text{(KM)}}(u) =
\begin{bmatrix}
\sigma_{\rho^{\text{(KM)}}}^2 f_{\rho}^{\text{(KM)}} + \int_\mathcal{I} \Sigma_{\rho^{\text{(KM)}} \bm{\beta}^{\text{(KM)}}} \bm{\phi}(u) f_{\beta}^{\text{(KM)}}(u) \,du \\
\bm{\phi}^\top(u) \Sigma_{\rho^{\text{(KM)}} \bm{\beta}^{\text{(KM)}}}^\top f_{\rho}^{\text{(KM)}} + \int_\mathcal{I} \mathcal{C}^{\text{(KM)}}(u, v) f_{\beta}^{\text{(KM)}}(v) \, dv
\end{bmatrix}.
\end{equation*}
\end{proof}

\begin{proof}[Proof of Theorem~\ref{th:2}]

Similar to \cite{CHERNOZHUKOV2006}, we adopt the empirical process notation as defined in \cite{van1996weak}. For the random variable set $\Z \equiv \left( Y, \bm{W} Y, \bm{\Lambda} \right)$, we define the following mappings:
\begin{align*}
f &\mapsto \mathbb{E}_{n} \left\{ f \left (\Z \right) \right\} \equiv n^{-1} \sum_{i=1}^{n} f \left( \Z_{i} \right), \\
f &\mapsto \mathbb{G}_{n} \left\{ f \left( \Z \right) \right\} \equiv n^{-1/2} \sum_{i=1}^{n} \left( f \left ( \Z_{i} \right) -\mathbb{E} \left \{ f \left ( \Z_{i} \right ) \right\} \right ).
\end{align*}
When $\widehat{f}$ is an estimated function, we denote:
\begin{align*}
\mathbb{G}_{n} \left\{ f \left ( \Z \right ) \right\} \Big|_{f = \widehat{f}} \equiv n^{-1/2} \sum_{i=1}^{n} \left( f \left( \Z_{i} \right) - \mathbb{E} \left\{ f \left( \Z_{i} \right) \right\} \right).
\end{align*}

Consider the parameter vector $\bm{\nu}_{\tau} \equiv \left( \bm{\beta}_{\tau}, \bm{\varsigma}_{\tau} \right)$ and define the transformation $\psi_{\tau}^{(-)} (c) \equiv -\psi_\tau (c)$. We introduce the following functions:
\begin{align*}
\widehat{f} \left( \Z, \rho_{\tau}, \bm{\nu}_{\tau} \right) &\equiv \psi_{\tau}^{(-)} \left( Y - \rho_{\tau} \bm{W} Y - \bm{\Xi} \bm{\beta}_{\tau} - \bm{\varsigma}_{\tau} \widehat{f}^* \right) \widehat{\bm{\Psi}}_{\tau}, \\
f \left( \Z, \rho_{\tau}, \bm{\nu}_{\tau} \right) &\equiv \psi_{\tau}^{(-)} \left( Y - \rho_{\tau} \bm{W} Y - \bm{\Xi} \bm{\beta}_{\tau} - \bm{\varsigma}_{\tau} f^* \right) \bm{\Psi}_{\tau},
\end{align*}
where $\widehat{\bm{\Psi}}_{\tau} = \widehat{T}(\bm{\Lambda}, \tau) (\widehat{f}^*, \bm{\Xi})$. Next, for a given function $\varphi_\tau(c)$, we define:
\begin{align*}
\widehat{g} \left( \Z, \rho_{\tau}, \bm{\nu}_{\tau} \right) &\equiv \varphi_{\tau} \left( Y - \rho_{\tau} \bm{W} Y -\bm{\Xi} \bm{\beta}_{\tau} - \bm{\varsigma}_{\tau} \widehat{f}^* \right) \widehat{T} (\bm{\Lambda}, \tau), \\
g \left( \Z, \rho_{\tau}, \bm{\nu}_{\tau} \right) &\equiv \varphi_{\tau} \left( Y - \rho_{\tau} \bm{W} Y - \bm{\Xi} \bm{\beta}_{\tau} - \bm{\varsigma}_{\tau} f^* \right) T (\bm{\Lambda}, \tau).
\end{align*}
Define the empirical and population expectations as $q_{n} \left(\rho_{\tau}, \bm{\nu}_{\tau} \right) \equiv \mathbb{E}_{n} \left\{ \widehat{g} \left( \Z, \rho_{\tau}, \bm{\nu}_{\tau} \right) \right\}$ and $q \left(\rho_{\tau}, \bm{\nu}_{\tau} \right) \equiv \mathbb{E} \left\{ g\left( \Z, \rho_{\tau}, \bm{\nu}_{\tau} \right) \right\}$, respectively. The optimal parameters are obtained by solving the minimization problems:
\begin{align*}
\widehat{\bm{\nu}}_{\tau} \left( \rho_{\tau} \right)  &\equiv \left\{ (\widehat{\bm{\beta}}^{\text{(CH)}}_{\tau} (\rho_{\tau}))^\top, \widehat{\bm{\varsigma}}_{\tau}(\rho_{\tau}) \right\} \equiv 
\underset{\bm{\nu}_{\tau} \in \mathbb{B} \times \mathbb{C}}{\arg\inf}~ q_{n} \left( \rho_{\tau}, \bm{\nu}_{\tau} \right), \\
\bm{\nu}_{\tau} \left( \rho_{\tau} \right) &\equiv \left\{ (\bm{\beta}_{\tau}(\rho_{\tau}))^\top, \bm{\varsigma}_{\tau}(\rho_{\tau}) \right\} \equiv \underset{\bm{\nu}_{\tau} \in \mathbb{B} \times \mathbb{C}}{\arg\inf}~ q\left( \rho_{\tau}, \bm{\nu}_{\tau} \right).
\end{align*}
Additionally, we define $\widehat{\rho}_{\tau}^{\text{(CH)}} \equiv \underset{\rho_{\tau} \in \mathbb{U}}{\arg\inf}~ \left\| \widehat{\bm{\varsigma}}_{\tau} \left( \rho_{\tau} \right) \right\|$, $\rho_{\tau}^{*} \equiv \underset{\rho_{\tau} \in \mathbb{U}}{\arg\inf}~ \left\| \bm{\varsigma}_{\tau} \left( \rho_{\tau} \right) \right\|$, $\widehat{\bm{\nu}}_{\tau} \equiv \left\{ (\widehat{\bm{\beta}}_{\tau}^{\text{(CH)}})^\top, \widehat{\bm{\varsigma}}_{\tau} \right\} \equiv \widehat{\bm{\nu}}_{\tau} \left( \widehat{\rho}_{\tau}^{\text{(CH)}} \right)$, and $\bm{\nu}_{\tau} \equiv \left( \bm{\beta}_{\tau}^\top, \bm{\varsigma}_{\tau} \right) \equiv \bm{\nu}_{\tau} \left( \rho_{\tau} \right)$.

We first establish the identifiability of the finite-dimensional parameters, namely, that $\left( \rho_{\tau}, \bm{\beta}_{\tau}^\top \right)^\top$ is uniquely determined by the limiting problem for each $\tau$. That is, we show that $\rho_{\tau}^{*} = \rho_{\tau}$ and $\bm{\beta}_{\tau} \left( \rho_{\tau}^* \right) = \bm{\beta}_{\tau}$. To this end, define:
\begin{align*}
\bigtriangledown \left( \rho_{\tau}, \bm{\beta}_{\tau}, \tau \right) &\equiv \mathbb{E} \left[ \varphi_{\tau} \left( Y - \rho_{\tau} \bm{W} Y - \bm{\Xi} \bm{\beta}_{\tau} \right) \bm{\Psi}_{\tau} \right], \\
\mathcal{J} \left( \rho_{\tau}, \bm{\beta}_{\tau}, \tau \right) &\equiv \frac{\partial }{\partial \left( \rho_{\tau}, \bm{\beta}_{\tau}^\top \right)} \mathbb{E} \left[ \varphi_{\tau} \left( Y - \rho_{\tau} \bm{W} Y - \bm{\Xi} \bm{\beta}_{\tau} \right) \bm{\Psi}_{\tau} \right].
\end{align*}
From conditions $C_8$ and $C_9$, the Jacobian matrix $\mathcal{J} \left( \rho_{\tau}, \bm{\beta}_{\tau}, \tau \right)$ is full rank and continuously varies with $\left( \rho_{\tau}, \bm{\beta}_{\tau}^{\top} \right)^\top$, uniformly over $\mathbb{U} \times \mathbb{B}$. Additionally, the image of $\mathbb{U} \times \mathbb{B}$ under the mapping $\left( \rho_{\tau}, \bm{\beta}_{\tau} \right)\mapsto \bigtriangledown \left( \rho_{\tau},\bm{\beta}_{\tau}, \tau \right)$ is assumed to be simply connected. 
By invoking Hadamard’s global univalence theorem for general metric spaces (e.g., Theorem 1.8 in \cite{ambrosetti1995primer}), as applied in \cite{chernozhukov2005iv}, it follows that the mapping $ \bigtriangledown \left( \cdot, \cdot, \tau \right)$ is a one-to-one homomorphism between $\left( \mathbb{U} \times \mathbb{B} \right)$ and its image $ \bigtriangledown \left( \mathbb{U}, \mathbb{B}, \tau \right)$. By \eqref{eq10_new}, the parameter vector $\left( \rho_{\tau}, \bm{\beta}_{\tau}^\top \right)^\top$ satisfies the equation:
\begin{equation}\label{eq4.14}
\mathbb{E} \left[ \varphi_{\tau} \left( Y - \rho_{\tau} \bm{W} Y - \bm{\Xi} \bm{\beta}_{\tau} - 0 f^* \right) \bm{\Psi}_{\tau} \right] = 0,
\end{equation}
which ensures its uniqueness in $\mathbb{U} \times \mathbb{B}$. This argument holds for every $\tau \in (0,1)$, thereby establishing the identifiability of the true parameters. Next, by assumption $A_8$ and the global convexity of $q \left( \rho_{\tau}, \bm{\nu}_{\tau} \right)$ in $\bm{\nu}_{\tau}$ for each $\tau$ and $\rho_{\tau}$, the parameter vector $\bm{\nu}_{\tau} \left( \rho_{\tau} \right)$ satisfies the sub-gradient condition:
\begin{equation}\label{eq4.15}
\mathbb{E} \left[ \varphi_{\tau} \left( Y - \rho_{\tau} \bm{W} Y - \bm{\Xi} \bm{\beta}_{\tau} \left( \rho_{\tau} \right) - \bm{\varsigma}_{\tau} \left( \rho_{\tau} \right) f^* \right) \bm{\Psi}_{\tau} \iota \right] \geq 0,
\end{equation}
for all $\iota$ such that $\bm{\nu}_{\tau} \left( \rho_{\tau} \right) + \iota \in \mathbb{B} \times \mathbb{C}$. If $\bm{\nu}_{\tau} \left( \rho_{\tau} \right)$ is in the interior of $\mathbb{B} \times \mathbb{C}$, then it uniquely solves the first-order condition:
\begin{equation}\label{eq4.16}
\mathbb{E} \left[ \varphi_{\tau} \left( Y - \rho_{\tau} \bm{W} Y - \bm{\Xi} \bm{\beta}_{\tau} \left( \rho_{\tau} \right) - \bm{\varsigma}_{\tau} \left( \rho_{\tau} \right) f^* \right) \bm{\Psi}_{\tau} \right] = 0.
\end{equation}
To determine $\rho_{\tau}^*$, we minimize $\left\| \bm{\varsigma}_{\tau } \left( \rho_{\tau} \right) \right\|$ over $\rho_{\tau}$ subject to \eqref{eq4.15}. From \eqref{eq4.14}, we see that $\rho_{\tau}^* = \rho_{\tau}$ leads to $\left\| \bm{\varsigma}_{\tau} \left( \rho_{\tau}^* \right) \right\| = 0$, thereby satisfying both \eqref{eq4.16} and \eqref{eq4.15} simultaneously. Since \eqref{eq4.14} ensures the uniqueness of the solution, it follows from \eqref{eq4.16} that $\bm{\beta}_{\tau} \left( \rho_{\tau}^* \right) = \bm{\beta}_{\tau}$.

Next, we establish the consistency of the estimators in the finite-dimensional space. By the bounded density condition in $C_9$, the function $q\left( \rho_{\tau}, \bm{\nu}_{\tau} \right)$ is continuous over the domain $\mathbb{U} \times \left( \mathbb{B} \times \mathbb{C} \right) \times (0,1)$. Furthermore, applying Lemma B.2 from \cite{CHERNOZHUKOV2006}, we obtain:
\begin{equation*}
\underset{\begin{subarray}{c} \left( \rho_{\tau}, \bm{\nu}_{\tau} \right) \in \mathbb{U} \times \left( \mathbb{B} \times \mathbb{C} \right) \times (0,1) \end{subarray}}{\sup}~ \left\| q_n \left( \rho_{\tau}, \bm{\nu}_{\tau} \right) - q \left( \rho_{\tau}, \bm{\nu}_{\tau} \right) \right\| \xrightarrow{p} 0.
\end{equation*}
By Lemma~B.1 in \cite{CHERNOZHUKOV2006}, this result implies the uniform convergence:
\begin{align*}
& \underset{\begin{subarray}{c} \rho_{\tau} \in \mathbb{U} \times (0,1) \end{subarray}}{\sup}~ \left\| \widehat{\bm{\nu}}_{\tau} \left( \rho_{\tau} \right) - \bm{\nu}_{\tau} \left( \rho_{\tau} \right) \right\| \xrightarrow{p} 0, \quad (*) \qquad \text{which further implies}, \\
& \underset{\begin{subarray}{c} \rho_{\tau} \in \mathbb{P} \times (0,1) \end{subarray}}{\sup}~ \left\| \left\| \widehat{\bm{\varsigma}}_{\tau} \left( \rho_{\tau} \right) \right\|_{A \left( \tau \right)} - \left\| \bm{\varsigma} _{\tau} \left( \rho_{\tau} \right) \right\|_{A \left( \tau \right)} \right\| \xrightarrow{p} 0.
\end{align*}
Applying Lemma B.1 from \cite{CHERNOZHUKOV2006} once more, we deduce:
\begin{equation*}
\underset{\begin{subarray}{c} \tau \in (0,1) \end{subarray}}{\sup}~ \left\| \widehat{\rho}_{\tau}^{(\text{CH})} -\rho_{\tau} \right\| \xrightarrow{p} 0.
\end{equation*}
Using $(*)$, this further implies $\underset{\begin{subarray}{c} \tau \in (0,1) \end{subarray}}{\sup}~ \left\| \widehat{\beta}_{\tau}^{(\text{CH})} - \beta_{\tau} \right\| \xrightarrow{p} 0, \quad \text{and} \quad \underset{\begin{subarray}{c} \tau \in (0,1) \end{subarray}}{\sup}~ \left\| \widehat{\bm{\varsigma}}_{\tau} \left( \widehat{\rho}_{\tau} \right) - 0 \right\| \xrightarrow{p} 0$. Finally, note that by the implicit function theorem, $\bm{\nu}_{\tau} \left( \rho_{\tau} \right)$ is continuous in both $\tau$ and $\rho_{\tau}$, and $\rho_{\tau}$ itself is continuous in $\tau$.

We now establish the asymptotic normality of the CH estimators. Consider a sequence of closed balls $\textbf{B}_{\gamma_n} \left( \rho_{\tau} \right)$ centered at $\rho_{\tau}$ for each $\tau$, where the radius $\gamma_n$ is independent of $\tau$ and satisfies $\gamma_n \to 0$ at a sufficiently slow rate. Let $\rho_n^{\left( \tau \right)}$ denote any point within $\textbf{B}_{\gamma_n} \left( \rho_{\tau} \right)$. By the computational properties of the standard quantile regression estimator $\widehat{\bm{\nu}}_{\tau} \left( \rho_n^{\left( \tau \right)} \right)$ \citep[see, e.g., Theorem 3.3 in][]{koenker1978}, we obtain:
\begin{equation}\label{eq4.17}
\mathcal{O}_p(1/\sqrt{n}) = \sqrt{n} \mathbb{E}_{n} \left\{ \widehat{f} \left( \Z, \rho_n^{(\tau)}, \widehat{\bm{\nu}}_{\tau}(\rho_n^{(\tau)}) \right) \right\}.
\end{equation}
By Lemma B.2 in \cite{CHERNOZHUKOV2006}, the following expansion holds for any sequence satisfying $\underset{\begin{subarray}{c}
\tau \in (0,1)
\end{subarray}}{\sup}~ \left\| \widehat{\rho}_n^{(\tau)} -\rho_{\tau} \right\|\xrightarrow{p} 0$. This allows us to express \eqref{eq4.17} as:
\begin{align}\label{eq4.18}
\begin{split}
\mathcal{O}_p(1/\sqrt{n}) &= \sqrt{n} \mathbb{E}_{n} \left\{ \widehat{f} \left( \Z, \rho_n^{(\tau)}, \widehat{\bm{\nu}}_{\tau}(\rho_n^{(\tau)}) \right) \right\} \\
&= 
\mathbb{G}_{n}\left\{ \widehat{f} \left( \Z, \rho_n^{(\tau)}, \widehat{\bm{\nu}}_{\tau}(\rho_n^{(\tau)}) \right) \right\} + \sqrt{n} \mathbb{E} \left\{ \widehat{f} \left( \Z, \rho_n^{(\tau)}, \widehat{\bm{\nu}}_{\tau}(\rho_n^{(\tau)}) \right) \right\} \\
&= 
\mathbb{G}_{n}\left\{ f \left( \Z, \rho_n^{(\tau)}, \bm{\nu}_{\tau}(\rho_n^{(\tau)}) \right) \right\} + o_p(1) + \sqrt{n} \mathbb{E} \left\{ \widehat{f} \left( \Z, \rho_n^{(\tau)}, \widehat{\bm{\nu}}_{\tau}(\rho_n^{(\tau)}) \right) \right\} \in \ell^{\infty} (0,1),
\end{split} 
\end{align}
where $\ell^{\infty} (0,1)$ denotes the space of bounded functions equipped with the supremum norm. Expanding the last line further, we obtain:
\begin{align}\label{eq4.19}
\begin{split}
\mathcal{O}_p(1/\sqrt{n}) =& \ \mathbb{G}_{n}\left\{ f \left( \Z, \rho_n^{(\tau)}, \bm{\nu}_{\tau}(\rho_n^{(\tau)}) \right) \right\}+ o_{p}(1) \\
&+\left( \mathcal{J}_{\bm{\nu}_{\tau}}(\tau) + o_p(1) \right) \sqrt{n} \left( \widehat{\bm{\nu}}_{\tau}(\rho_n^{(\tau)}) - \bm{\nu}_{\tau} \right) \\
&+ \left( \mathcal{J}_{\rho_{\tau}}(\tau) + o_p(1) \right) \sqrt{n} \left( \rho_n^{(\tau)} - \rho_{\tau} \right) \in \ell^{\infty} (0,1),
\end{split} 
\end{align}
where the Jacobian matrices are defined as:
\begin{align*}
\mathcal{J}_{\bm{\nu}_{\tau}}(\tau) &= \frac{\partial }{\partial \left( \bm{\beta}_{\tau}^\top, \bm{\varsigma}_{\tau} \right)} \mathbb{E} \left\{ \varphi_{\tau} \left( Y - \rho_{\tau} \bm{W} Y - \bm{\Xi} \bm{\beta}_{\tau} - \bm{\varsigma}_{\tau} \vartheta \right) \bm{\Psi}_{\tau} \right\}_{\left( \bm{\varsigma}_{\tau}, \bm{\beta}_{\tau} \right) = \left( 0, \bm{\beta}_{\tau} \right)}, \\
\mathcal{J}_{\rho_{\tau}}(\tau) &= \frac{\partial }{\partial \left( \rho_{\tau} \right)} \mathbb{E} \left\{ \varphi_{\tau} \left( Y - \rho_{\tau} \bm{W} Y - \bm{\Xi} \bm{\beta}_{\tau} \right) \bm{\Psi}_{\tau} \right\}_{\left( \rho_{\tau} \right) = \left( \rho_{\tau} \right )}.
\end{align*}
In other words, for any sequence satisfying $\underset{\begin{subarray}{c} \tau \in (0,1) \end{subarray}}{\sup}~ \left\| \rho_n^{\tau} -\rho_{\tau} \right\| \xrightarrow{p} 0$, we derive the asymptotic expansion:
\begin{align*}
\begin{split}
\sqrt{n} \left( \widehat{\bm{\nu}}_{\tau}(\rho_n^{(\tau)}) - \bm{\nu}_{\tau} \right) =& \ -\mathcal{J}_{\bm{\nu}_{\tau}}^{-1}(\tau) \mathbb{G}_{n} \left\{ f \left( \Z, \rho_{\tau}, \bm{\nu}_{\tau} \right) \right\} - \mathcal{J}_{\bm{\nu}_{\tau}}^{-1}(\tau) \mathcal{J}_{\rho_{\tau}}(\tau) \left\{ 1+o_p(1) \right\} \\
& \times \sqrt{n} \left( \rho_n^{(\tau)} - \rho_{\tau} \right) + o_p(1) \in \ell^{\infty} (0,1),
\end{split}
\end{align*}
which implies:
\begin{align*}
\begin{split}
\sqrt{n} \left( \widehat{\bm{\varsigma}}_{\tau}(\rho_n^{(\tau)}) - 0 \right) =& -\overline{\mathcal{J}}_{\bm{\varsigma}_{\tau}}^{-1}(\tau) \mathbb{G}_{n} \left\{ f \left( \Z, \rho_{\tau}, \bm{\nu}_{\tau} \right) \right\} - \overline{\mathcal{J}}_{\bm{\varsigma}_{\tau}}(\tau) \mathcal{J}_{\rho_{\tau}}(\tau) \left\{1 + o_p(1) \right\} \\
& \times \sqrt{n} \left( \rho_n^{(\tau)} - \rho_{\tau} \right) + o_p(1) \in \ell^{\infty} (0,1),
\end{split}
\end{align*}
where $\left\{ \overline{\mathcal{J}}_{\bm{\beta}_{\tau}}^{\top}(\tau) : \overline{\mathcal{J}}_{\bm{\varsigma}_{\tau}}(\tau) \right\}^\top$ represents the conformable partition of $\mathcal{J}_{\bm{\nu}_{\tau}}^{-1}(\tau)$.

By the consistency result and with inner probability tending to 1, we have:
\begin{equation*}
\widehat{\rho}_{\tau}^{\text{(CH)}} = \underset{\begin{subarray}{c}
\rho_n^{\tau} \in \bm{B}_{n}(\rho_{\tau})
\end{subarray}}{\arg\inf}~
\left\| \widehat{\bm{\varsigma}}_{\tau} \left( \rho_{n}^{\tau} \right) \right\|_{A(\tau)} \quad 
\text{for all} \quad \tau \in (0,1).
\end{equation*}
Applying Lemma B.2 from \cite{CHERNOZHUKOV2006}, we obtain the stochastic bound $\mathbb{G}_{n} \left \{ f \left( \Z, \rho_{\tau}, \bm{\nu}_{\tau} \right) \right\} = \mathcal{O}_p(1)$. Thus, we deduce:
\begin{equation*}
\sqrt{n} \left\| \widehat{\bm{\varsigma}}_{\tau}(\rho_n^{(\tau)}) \right\|_{A(\tau)} = 
\left\| \mathcal{O}_p(1) - \overline{J}_{\bm{\varsigma}_{\tau}}(\tau) \mathcal{J}_{\rho_{\tau}}(\tau) \left\{ 1 + o_p(1) \right\} \sqrt{n} \left( \rho_n^{(\tau)} - \rho_{\tau} \right) \right\|_{A(\tau)} \in \ell^{\infty} (0,1).
\end{equation*}
Since $\overline{\mathcal{J}}_{\bm{\varsigma}_{\tau}}(\tau) \mathcal{J}_{\rho_{\tau}}(\tau)$ and $A(\tau)$ are of full rank uniformly in $\tau$, it follows that $\sqrt{n} \left( \widehat{\rho}_{\tau}^{\text{(CH)}} - \rho_{\tau} \right) = \mathcal{O}_p(1) \in \ell^{\infty} (0,1)$. Using arguments similar to those in Lemma B.1 of \cite{CHERNOZHUKOV2006}, we obtain:
\begin{equation*}
\sqrt{n} \left( \widehat{\rho}_{\tau}^{\text{(CH)}} - \rho_{\tau} \right) = \underset{\begin{subarray}{c}
\mu \in \bm{R}^{\ell}
\end{subarray}}{\arg\inf} \left\| -\overline{\mathcal{J}}_{\bm{\varsigma}_{\tau}}(\tau)\mathbb{G}_{n} \left\{ f \left( \Z, \rho_{\tau}, \bm{\nu}_{\tau} \right) \right\}- \overline{\mathcal{J}}_{\bm{\nu}_{\tau}}(\tau) \mathcal{J}_{\rho_{\tau}}(\tau) \mu \right\|_{A(\tau)} + o_p(1) \in \ell^{\infty} (0,1).
\end{equation*}
Thus, in the space $\ell^{\infty} (0,1)$, we conclude:
\begin{align*}
\begin{split}
\sqrt{n} \left( \widehat{\rho}_{\tau}^{\text{(CH)}} - \rho_{\tau} \right) =& -\left( \mathcal{J}_{\rho_{\tau}}^{\top}(\tau) \overline{\mathcal{J}}_{\bm{\varsigma}_{\tau}}^{\top}(\tau) A (\tau) \overline{\mathcal{J}}_{\bm{\varsigma}_{\tau}}(\tau) \mathcal{J}_{\rho_{\tau}}(\tau) \right)^{-1} 
\left( \mathcal{J}_{\rho_{\tau}}^{\top}(\tau) \overline{\mathcal{J}}_{\bm{\varsigma}_{\tau}}^{\top}(\tau) A(\tau) \overline{\mathcal{J}}_{\bm{\varsigma}_{\tau}}(\tau) \right) \\
&\times \mathbb{G}_{n} \left\{ f \left( \Z, \rho_{\tau}, \nu_{\tau} \right) \right\} + o_p(1) = \mathcal{O}_p(1).
\end{split} 
\end{align*}
Similarly, for the parameter vector $\bm{\nu}_{\tau}$:
\begin{align*}\small
\begin{split}
\sqrt{n} \left( \widehat{\bm{\nu}}_{\tau}(\widehat{\rho}_{\tau}^{\text{(CH)}}) - \bm{\nu}_{\tau} \right) =& 
- \mathcal{J}_{\bm{\nu}_{\tau}}^{-1}(\tau) \left[ \mathit{I} - \mathcal{J}_{\rho_{\tau}}(\tau) \left\{ \mathcal{J}_{\rho_{\tau}}^{\top}(\tau) \overline{\mathcal{J}}_{\bm{\varsigma}_{\tau}}^{\top}(\tau)A(\tau) \overline{\mathcal{J}}_{\bm{\varsigma}_{\tau}}(\tau) \mathcal{J}_{\rho_{\tau}}(\tau) \right\}^{-1} \mathcal{J}_{\rho_{\tau}}^{\top}(\tau) \overline{\mathcal{J}}_{\bm{\varsigma}_{\tau}}^{\top}(\tau) A(\tau) \overline{\mathcal{J}}_{\bm{\varsigma}_{\tau}}(\tau) \right ] \\
&\times \mathbb{G}_{n} \left\{ f \left( \Z, \rho_{\tau}, \bm{\nu}_{\tau} \right) \right\} + o_p(1) = \mathcal{O}_p(1).
\end{split}
\end{align*}

Due to the invertibility of $\mathcal{J}_{\rho_{\tau}}(\tau) \overline{\mathcal{J}}_{\bm{\varsigma}_{\tau}}(\tau)$, we obtain:
\begin{align*}
\begin{split}
\sqrt{n} \left\{ \widehat{\bm{\varsigma}}_{\tau}(\widehat{\rho}_{\tau}^{\text{(CH)}}) - 0 \right\} =& 
-\overline{\mathcal{J}}_{\bm{\varsigma}_{\tau}}(\tau) \left[ \mathit{I} - \mathcal{J}_{\rho_{\tau}}(\tau) \left\{ \mathcal{J}_{\rho_{\tau}}^{\top}(\tau) \overline{\mathcal{J}}_{\bm{\varsigma}_{\tau}}^{\top}(\tau) \right\}^{-1} \overline{\mathcal{J}}_{\bm{\varsigma}_{\tau}}(\tau) 
\right] \\
&\times \mathbb{G}_{n} \left\{ f \left( \Z, \rho_{\tau}, \bm{\nu}_{\tau} \right) \right\}+ o_p(1) \\
=&\ 0 \times \mathcal{O}_p(1) + o_p(1) \in \ell^{\infty} (0,1).
\end{split}
\end{align*}
Rather than explicitly performing algebraic simplifications, we utilize this result by substituting $\left\{ \rho_{n}^{(\tau)}, \widehat{\bm{\nu}}_{\tau}(\rho_{n}^{(\tau)}) \right\} = \left(\widehat{\rho}_{\tau}, \widehat{\bm{\nu}}_{\tau}  \right) = \left\{ \widehat{\rho}_{\tau}^{\text{(CH)}},\widehat{\bm{\beta}}_{\tau}^{\text{(CH)}}, 0 + o_p(1/\sqrt{n}) \right\}$ back into the expansion~\eqref{eq4.19}, yielding:
\begin{equation*}
-\mathbb{G}_{n} \left\{ f \left( \Z, \rho_{\tau}, \bm{\nu}_{\tau} \right) \right\} = \mathcal{J}(\tau) \sqrt{n} (\widehat{\bm{\alpha}}_{\tau}^{\text{(CH)}} - \bm{\alpha}_\tau) + o_p(1) \in \ell^{\infty} (0,1).
\end{equation*}
By Lemma B.2 in \cite{CHERNOZHUKOV2006}, we obtain $\mathbb{G}_{n} \left\{ f \left( \Z, \rho_{\tau}, \bm{\nu}_{\tau} \right) \right\} \Rightarrow  \mathbb{G}(\tau) \in \ell^{\infty} (0,1)$, where $\mathbb{G}(\tau)$ is a Gaussian process with covariance function $\mathcal{V}(\tau, \tau^\prime) = \left\{ \min \left( \tau, \tau^\prime \right) - \tau \tau^\prime \right\} \mathbb{E} \left\{ \bm{\Psi}_{\tau} \bm{\Psi}_{\tau^\prime}^{\top} \right\}$. This result leads to the desired asymptotic distribution:
\begin{equation*}
\sqrt{n} (\widehat{\bm{\alpha}}_{\tau}^{\text{(CH)}} - \bm{\alpha}_\tau) \overset{d}{\rightarrow} \mathcal{N} ( 0, \bm{\Sigma}_{\bm{\alpha}^{\text{(CH)}}} ) \in \ell^{\infty} (0,1),
\end{equation*}
where $\bm{\Sigma}_{\bm{\alpha}^{\text{(CH)}}} = \mathcal{J}^{-1}(\tau) \mathcal{V}(\tau \tau^\prime) (\mathcal{J}^{-1}(\tau^\prime))^\top$.

Analogous to the proof of Theorem~\ref{th:1}, we extend the results to the functional space. The functional regression coefficient estimator, $\widehat{\beta}_\tau^{\text{(CH)}}(u)$, can be represented as $\widehat{\beta}_\tau^{\text{(CH)}}(u) = \bm{\phi}^\top(u) \widehat{\bm{\beta}}_\tau^{\text{(CH)}}$. The asymptotic behavior of the scalar estimator $\widehat{\rho}_\tau^{\text{(CH)}}$ follows a normal distribution with asymptotic variance $\sigma^2_{\rho^{\text{(CH)}}}$, given by the $(1,1)$-th entry of $\bm{\Sigma}_{\bm{\alpha}^{\text{(CH)}}}$. Furthermore, the process $\sqrt{n} \{\widehat{\beta}^{\text{(CH)}}_\tau(u) - \beta_\tau(u)\}$ converges weakly to a Gaussian process with covariance function $\mathcal{C}^{\text{(CH)}}(u,v) = \bm{\phi}^\top(u) \bm{\Sigma}_{\bm{\beta}^{\text{(CH)}}} \bm{\phi}^\top(v)$, where $\bm{\Sigma}_{\bm{\beta}^{\text{(CH)}}}$ is the $M \times M$ submatrix of $\bm{\Sigma}_{\bm{\alpha}^{\text{(CH)}}}$ corresponding to $\bm{\beta}_\tau$. The covariance between the normalized deviations $\sqrt{n} (\widehat{\rho}_\tau^{\text{(CH)}} - \rho_\tau)$ and $\sqrt{n} \{\widehat{\beta}^{\text{(CH)}}_\tau(u) - \beta_\tau(u)\}$ is given by:
\begin{equation*}
\operatorname{Cov} \Big[ \sqrt{n} (\widehat{\rho}_\tau^{\text{(CH)}} - \rho_\tau), \sqrt{n} \{\widehat{\beta}^{\text{(CH)}}_\tau(u) - \beta_\tau(u)\} \Big] = \bm{\Sigma}_{\rho^{\text{(CH)}} \bm{\beta}^{\text{(CH)}}} \bm{\phi}(u),
\end{equation*}
where $\bm{\Sigma}_{\rho^{\text{(CH)}} \bm{\beta}^{\text{(CH)}}}$ represents the $1 \times M$ block of $\bm{\Sigma}_{\bm{\alpha}^{\text{(CH)}}}$ associated with the covariance between $\rho_\tau$ and $\bm{\beta}_\tau$.

By assuming conditions $C_1$-$C_5$ and $C_8$-$C_{10}$, the estimators $\widehat{\rho}_\tau^{\text{(CH)}}$ and $\widehat{\beta}_\tau^{\text{(CH)}}(u)$ jointly converge in distribution to a Gaussian process:
\begin{equation*}
\sqrt{n} \begin{bmatrix} \widehat{\rho}_\tau^{\text{(CH)}} - \rho_\tau \\ \widehat{\beta}_\tau^{\text{(CH)}}(u) - \beta_\tau(u) \end{bmatrix} \xrightarrow{d} \mathcal{GP} \left( 0, \Sigma_{\bm{\theta}^{\text{(CH)}}} \right),
\end{equation*}
where the covariance operator $\Sigma_{\bm{\theta}^{\text{(CH)}}}$ is defined for functions $f^{\text{(CH)}}(u) = \{f_{\rho}^{\text{(CH)}}, f_{\beta}^{\text{(CH)}}(u)\}$ as:
\begin{equation*}
\Sigma_{\bm{\theta}^{\text{(CH)}}} f^{\text{(CH)}}(u) =
\begin{bmatrix}
\sigma_{\rho^{\text{(CH)}}}^2 f_{\rho}^{\text{(CH)}} + \int_\mathcal{I} \Sigma_{\rho^{\text{(CH)}} \bm{\beta}^{\text{(CH)}}} \bm{\phi}(u) f_{\beta}^{\text{(CH)}}(u) \,du \\
\bm{\phi}^\top(u) \Sigma_{\rho^{\text{(CH)}} \bm{\beta}^{\text{(CH)}}}^\top f_{\rho}^{\text{(CH)}} + \int_\mathcal{I} \mathcal{C}^{\text{(CH)}}(u, v) f_{\beta}^{\text{(CH)}}(v) \, dv
\end{bmatrix}.
\end{equation*}
\end{proof}

\section*{Acknowledgments}

The authors acknowledge the insightful comments from the participants of the 31\textsuperscript{st} conference of the International Environmetrics Society (TIES2024). This research was partially supported by the Scientific and Technological Research Council of Turkey (TUBITAK) (grant no. 124F096) and the Australian Research Council Future Fellowship (grant no. FT240100338). 

%\newpage
\bibliographystyle{agsm}
\bibliography{SSoFQRM.bib}

\end{document}